\documentclass[twocolumn,showpacs,nopreprintnumbers,aps,prd, letterpaper,groupeaddress,nofootinbib,tightenlines,floats,floatfix,showkeys,superscriptaddress]{revtex4-2}

\usepackage{graphicx}
\usepackage{float}
\usepackage{dcolumn} 
\usepackage{bm}
\usepackage{hyperref}
\usepackage[english]{babel}
\usepackage[utf8]{inputenc}
\usepackage[usenames,dvipsnames]{color}
\usepackage{amsmath,amssymb}
\usepackage{physics}
\usepackage{slashed}
\usepackage{appendix}
\usepackage{soul}
\usepackage{ulem}
\usepackage{booktabs}
\usepackage{multirow}
\usepackage[dvipsnames]{xcolor}
\usepackage{orcidlink}
\usepackage[caption=false]{subfig}
\usepackage{color}
\hypersetup{
    colorlinks=true,
    linkcolor=blue,
    urlcolor=blue,
    filecolor=magenta,      
    citecolor=red
}

\begin{document}

\title{Scalar field dark matter with two components:\\Combined approach from particle physics and cosmology}

\author{Er\'endira Guti\'errez-Luna \orcidlink{0000-0002-1415-2799}}
\affiliation{Instituto de F\'isica, Universidad Nacional Aut\'onoma de M\'exico, A.P. 20-364, 01000, M\'exico D.F., M\'exico}
\email{lgutierrez@estudiantes.fisica.unam.mx} 

\author{Belen Carvente \orcidlink{0000-0002-7050-7769}}
\affiliation{Instituto de Ciencias Nucleares, Universidad Nacional Aut\'{o}noma de M\'{e}xico, 
Circuito Exterior C.U., A.P. 70-543, M\'exico D.F. 04510, M\'{e}xico}

\author{V\'ictor Jaramillo\orcidlink{0000-0002-3235-4562}}
\affiliation{Instituto de Ciencias Nucleares, Universidad Nacional Aut\'{o}noma de M\'{e}xico, 
Circuito Exterior C.U., A.P. 70-543, M\'exico D.F. 04510, M\'{e}xico}

\author{Juan Barranco\orcidlink{0000-0002-9511-6772}}
\affiliation{Departamento de F\'isica, Divisi\'on de Ciencias e Ingenier\'ias,
Campus Le\'on, Universidad de Guanajuato, Le\'on 37150, M\'exico}

\author{Celia Escamilla-Rivera\orcidlink{0000-0002-8929-250X}}
\affiliation{Instituto de Ciencias Nucleares, Universidad Nacional Aut\'{o}noma de M\'{e}xico, 
Circuito Exterior C.U., A.P. 70-543, M\'exico D.F. 04510, M\'{e}xico}

\author{Catalina Espinoza\orcidlink{0000-0001-7074-1726}}
\affiliation{C\'{a}tedras Conacyt, Instituto de F\'isica, Universidad Nacional Aut\'onoma de M\'exico, A.P. 20-364, 01000, M\'exico D.F., M\'exico}

\author{Myriam Mondrag\'on\orcidlink{0000-0002-5081-9474}}
\affiliation{Instituto de F\'isica, Universidad Nacional Aut\'onoma de M\'exico, A.P. 20-364, 01000, M\'exico D.F., M\'exico}

\author{Dar\'io N\'u\~nez\orcidlink{0000-0003-0295-0053}}
\affiliation{Instituto de Ciencias Nucleares, Universidad Nacional Aut\'{o}noma de M\'{e}xico, 
Circuito Exterior C.U., A.P. 70-543, M\'exico D.F. 04510, M\'{e}xico}

\begin{abstract}
In this work we explore the possibility of incorporating particle physics motivated scalar fields to the dark matter cosmological model. In this landscape, we consider the classical complex scalar field  
in a certain region in the parameter space of the model that increases the number of neutrino species $N_{\mathrm{eff}}$, in order to 
be consistent with the observed abundance of light elements produced at Big Bang Nucleosynthesis (BBN). We perform analyses using one and two scalar fields. We examine the difference between these models and the priors considered at the edges of the cosmic ladder, this with the purpose of studying the impact of such models on the Hubble cosmic flow. 
In the two scalar field models we explore the possibility of combining an axion and a Higgs-like field as well as a Higgs-like field and the classical field, we show that in the first case there is no set of parameters that allows us to be consistent with $N_\mathrm{eff}$, while in the second case a strong restriction to the set of parameters is obtained. This last restriction is given in terms of a maximum bound of the fraction of Higgs-like field that can be incorporated together with the classical field. Our results could be relevant in the direct dark matter detection programs.

\end{abstract}

\maketitle

\section{Introduction}
\label{sec:level1}

Over the years there has been a remarkable development regarding the studies on the dark matter (DM) component of the Universe. The cosmological observations have been more precise and left no doubt that, within our present understanding of the fundamental interactions, there is in the Universe a component $6.5$ larger than the amount of observed baryonic matter \cite{Planck:2018vyg} that, except for the gravitational one, has a very small interaction with the observable matter and very low electromagnetic emission \cite{Dodelson:2003ft}. Actually, projects as DAMA, CRESST, IceCube, and PandaX \cite{2013EPJC...73.2648B, 2015arXiv150400820D, 2016EPJC...76...25A, 2017PhRvL.119r1302C,2011ICRC....5..141D, 2015JCAP...09..008F,2020NuPhA.99621712Y}, aimed to the detection of a dark matter particle, have not been able to obtain any detection. We must face the possibility that the interaction of the baryonic or leptonic matter and dark matter, besides the gravitational, might be zero. 

The proposal of modeling the dark matter as a scalar field endowed with a scalar potential of the form 
\begin{equation}
\label{eq:potential1}
V(|\phi|)=\mu^2\,|\phi|^2 + \sigma^2\,|\phi|^4, 
\end{equation}
has grown since the early work discussed in \cite{Matos:1998vk} (see also references therein), where it was shown that a real scalar field with a very small parameter, $\mu$, and no quartic term, $\sigma=0$, could describe a galactic halo and avoid some problems of the standard weakly interacting massive particle (WIMP) model, like the super abundance of satellites predicted by cosmological simulations. Nowadays the proposal has received serious consideration by the community, see for example \cite{Hui:2016ltb,Urena-Lopez:2019kud,Suarez:2013iw}, as several of the benchmarks for a cosmological model have been successfully performed by such a model, called an ultralight scalar field, as the parameter $\mu$ can be related to the mass of the boson particle, $m_\phi$, with the expression $\mu=\frac{m_\phi\,c}{\hbar}$, where $c$ stands for the speed of light in a vacuum and $\hbar$ for the reduced Planck's constant, such a model has also been called {\it fuzzy} dark matter. Using this model, it has been possible to reproduce the large scale fiber structure observed in the Universe \cite{Schive:2014dra,Mocz:2019uyd}, as well as the observed harmonic structure of the perturbations \cite{Cembranos:2015oya, Amendola:1999dr}; the galactic halos and the observed rotational velocity profiles in the galaxies has important developments within this model \cite{Bernal:2003lwr}; the quartic parameter, $\sigma$, is interpreted as describing the self-interaction of the field. It is interesting that if the units of the scalar field are absorbed in a constant in the Lagrangian, then one can consider a scalar field described by a unitless function
and both parameters $\mu$ and $\sigma$ have units of inverse of distance; and as long as the scalar field satisfies the Klein-Gordon equation, which is a wavelike description, one can then interpret the parameters as the De Broglie wavelength of the scalar field, $\lambda_\phi$,  and the gravitational equilibrium scale \cite{Li:2013nal}, allowing us to call such models an ultralong wavelength scalar field.
In this paper, however, regarding the scalar field, we will use the usual unit conventions 
both to make smoother the passage from the quantum field theory (QFT) to the classical one and to make the cosmological analysis in the usual way. 

Models considering a complex scalar field for describing dark matter (and even dark energy, see \cite{Carvente:2020aae}) have also been considered and have proved to 
give a consistent description of the Fridman homogeneous Universe \cite{Li:2013nal} (we are using the direct transliteration from the Russian name), with a scalar potential as in Eq.~(\ref{eq:potential1}), showing that the $\mu$ parameter needs not to be very small, it is enough to demand that $m_\phi>10^{-21}\,{\rm eV}/c^2$, due to the presence of the quartic term $\sigma$ \cite{Li:2013nal,Suarez:2016eez}, which is strongly constrained in terms of the combination $\sigma^2/\mu^4$, proportional to the gravitational length scale that $\sigma$ defines, which turns out to be ultralong, of the order of kiloparsecs \cite{Li:2013nal, Suarez:2016eez}. These scalar fields are considered completely noninteracting with other types of matter, and we will call them {\it classical scalar fields}. There is a growing conviction not only that scalar fields are very plausible candidates to describe the dark matter present in the Universe, but that objects described by such scalar fields very plausibly exist in nature. Models considering a real scalar field have also been considered in large scale cosmology, see \cite{Torres:2014bpa, Chakraborty:2021vcr} for instance, but they induce the wave oscillation to the spacetime structure, as in the case of the compact objects they form, called "oscillatons" \cite{Alcubierre:2003sx}, and such oscillations in the scale factor could impose strong constraints on the value of the real scalar field parameters. In this work, we will consider complex scalar fields that do not present such oscillations in the spacetime geometry. 

The cosmological 
is a serious alternative to the standard cold dark matter(CDM) model, where the dark matter is treated as a pressureless fluid of WIMPs \cite{Dodelson:2003ft}.

Such classical scalar fields, as long as they are considered completely noninteracting with any other field or particle, save via the gravitational interaction, of course, can be incorporated to the Standard Model (SM) of particle physics, along with the other dark matter scalar field models
 such as the axion or the Higgs-like, which, in the classical limit, can be also described as a complex scalar field with a scalar potential as given above but with different values of the parameters $\mu$ and $\sigma$. One can naturally ask how much of 
 the axion or Higgs-like fields can be present along with the classical one, maintaining the general properties that make the single classical field appealing.

The main goal of this paper is to shed light on this question and, therefore, its consequences at cosmological scales. As mentioned above, matter classically described by a complex scalar field with very large values of the mass parameter, of the order of ${\rm eV}, \, {\rm keV}$, or even hundreds of ${\rm GeV}$ \cite{ParticleDataGroup:2020ssz, Espinoza:2018itz, OSQAR:2015qdv, Ehret:2010mh, GAMBIT:2018eea, LHCDarkMatterWorkingGroup:2018ufk, Graham:2013gfa, Robilliard:2007bq, Cameron:1993mr, GammeVT-969:2007pci, OSQAR:2007oyv, PVLAS:2007wzd, Marsh:2015xka} (corresponds to a wavelength of $10^{-7}, 10^{-10},10^{-18} {\rm m}$ respectively) could exist in nature as constituents of dark matter and be part of the general content of the Universe, but how much of a Higgs-like or an axion field could be considered as a component of the evaluated dark matter? Indeed, there is no reason to consider that the dark matter sector should be described by a single type of matter; we could have the classical as well as other scalar fields included in the computation of the dark matter density. Such considerations could reduce some pressure to the groups in the direct search of dark matter, mentioned above, as long as heavier scalar fields, which are the ones usually searched for, might not be the total of the dark matter density, and thus the detection probability is reduced by in a significant amount. In the present work we will consider that the dark matter sector of the Universe is described by two complex scalar fields. 

The passage from a particle physics model with foundations in a quantum field theory to a semiclassical description is often assumed obvious in the literature. For the sake of clarity of the expositions in the following sections we give a brief argument on this matter. In order to study the cosmological implications of such quantum models,
a clean path to follow 
is to first take the classical limit in order to be able to embed the corresponding model’s classical fields into a gravitational action, by coupling them in a minimal way to the gravitational field. Next we simply assume that these fields obey semiclassical equations of motion, for the case of the scalar fields considered in this work these would be the Klein-Gordon equation. Finally, we study the cosmological implications of the resulting setups. Of course, in practice it is sufficient to identify the field content of the quantum model and pass directly to the semiclassical equations, but we feel it is important to give a slightly more formal argumentation for this step (we also expand briefly on the classical limit of a QFT in section \ref{sec:transition}).

In section \ref{sec:level2} we part from the 
particle physics and discuss how the axion-like or the Higgs-like particles can be described by a complex scalar field with a scalar potential of the type described above in the semi-classical limit, discussing also the range of values of the parameters of the potential. In section \ref{sec:two_scalar} we describe the homogeneous Fridman model with two such scalar fields and the integration procedure of the field equations. This approach based on \cite{Li:2013nal} will be generalized to solve for the negative self-interaction fully relativistic scalar field. Then, in section \ref{sec:evol} we present the evolution of certain reference cases for the classical, axion and Higgs-like scalar field models. 

After representative single scalar field cases are presented, we will show some of the solutions for the combinations classical+axion and axion+Higgs, where a new parameter $\eta$ will enter since we need to fix the relative fraction of energy density of each of the fields at late times with respect to the total dark matter.

In section \ref{sec:obsevational} we present the results of this work, namely, the cosmological effects of considering as part of the matter content the scalar fields described in the previous sections as well as specific cosmological constraints for the two scalar field parameters. In subsection \ref{sec:ladder} we examine the variability between the two scalar field models and the priors considered at the edges of the cosmic ladder, namely the $H_0$ value at early and late times \cite{DiValentino:2020zio}, obtaining that they have a clearly different behavior depending on the combination of the two scalar fields taken into account. Next, in subsection \ref{sec:neff} we present a discussion on how the number of neutrino species $N_{\mathrm{eff}}$ together with the requirement of the scalar field to behave as matter at the matter-radiation equality sets a very restrictive condition on the parameters of the model. 

In the concluding section \ref{sec:conclusions} we summarize the more relevant results and a discussion on the implications on direct dark matter detection programs as well as in the cosmological and astrophysical dark matter research.

\section{\label{sec:level2} Scalar field dark matter in particle physics}

The Standard Model of particle physics  describes the phenomena observed so far in elementary particle physics with very good precision. However, this successful model  can only help us to understand about $5\%$ of the total matter in the Universe. It is assumed to be the low energy limit 
of a more fundamental theory, and  must be extended to explain other phenomena like neutrino masses, matter-antimatter asymmetry and even dark matter.

The remaining components in the Universe, called dark matter and dark energy, which make up about $27\%$ and $67\%$ of the total matter in the Universe, respectively \cite{Planck:2018vyg}, do not find an explanation in the framework of the SM but their existence is inferred from its gravitational effects in the astrophysical observations \cite{Planck:2019evm, Planck:2018lbu}. Beyond the facts relating to the temperature and longevity of the dark matter, we have very little information about its nature and properties. 
In addition, the lack of experimental evidence in the search for the most popular candidates such as WIMP, sterile neutrinos or dark photons, makes evident the need for new models and search techniques for possible DM candidates. The dark matter may well consist of one or more types of fundamental particles. The simplest fundamental particle is a scalar field (zero spin particle).

Among the most common candidates to scalar field dark matter (SFDM) in particle physics are axions, axion-like  and Higgs-like particles. In particular, we are interested in a model that includes two scalar fields. We will consider here  one of the candidates to come from an inert scalar $SU(2)$ doublet, {\it i.e. Higgs-like}, motivated by some  extensions of the SM, where this proposal has been successful \cite{LopezHonorez:2006gr, Cao:2007rm, Majumdar:2006nt}. A second candidate may be an axion or axion-like particle coming from particle physics or cosmology \cite{Marsh:2015xka, chadhaday2021axion}, and both will be worked along with the classical complex scalar field mentioned in the introduction. 

\subsection{Axion and axion-like particles} \label{axion-likesubsec}
The word ``axion'' can take on a variety of meanings. The first time was used to name the particle associated to the Peccei-Quinn (PQ) mechanism for preserving Charge-Parity (CP) symmetry in the strong interactions \cite{Peccei:1977hh, Weinberg:1977ma, Wilczek:1977pj}.
Legend says that F. Wilczek, who was looking for a name to describe a new pseudo Goldtone boson, while washing clothes, looked at the name of the detergent he was using, {\it axion} and decided to use that name for the new particle, since he expected it would clean up the problem of QCD with CP symmetry.

Parity (P) is the space reflection operator, i.e. inverts the spatial coordinates, $P: \Vec{x} \rightarrow -\vec{x}$ and the charge conjugation operator (C), changes particles into antiparticles without affecting their momenta or spin \cite{Langacker:2010zza}. In a decay, the combined transformation CP changes particles to antiparticles and the sense of longitudinal polarization is reversed. If, the rate for one decay and its conjugate are the same, then the CP symmetry is conserved. 

In quantum field theory, the term ``axion'' applies to any pseudoscalar Goldstone boson of the spontaneous breaking of one global chiral symmetry that is broken at some scale $f_a$. Such particles need not solve the strong CP problem or couple to gluons \cite{chadhaday2021axion}. This means their mass could take any value and be very weakly coupled which makes them difficult to detect experimentally. These Goldstone bosons that do not acquire a mass from radiative corrections of Quantum Chromodynamics (QCD) are also called axion-like-particles (ALPs).

In string theory the term ``axion'' can refer either to matter fields, or to pseudoscalar fields associated to the geometry of compact spatial dimensions \cite{Marsh:2015xka}. From now on, we will use the word ``axion'' to refer to a pseudoscalar field in any of the theories mentioned above.

The axion acquires mass from QCD chiral symmetry breaking, and can be calculated in chiral perturbation theory \cite{Weinberg:1977ma, Marsh:2015xka},
\begin{equation} \label{axionmass}
 m_a \approx 6 \mu \text{eV}  \left( \frac{10^{12} \text{GeV}}{f_a} \right) .
\end{equation}
This expression is a largely model-independent statement. The axion decay constant $f_a$ is related to vacuum expectation value $v_a$, that breaks the Peccei-Quinn symmetry $f_a = v_a/N_{DW}$. $N_{DW}$ is an integer that characterizes the vacuum of axion models called a \textit{color anomaly}, also known as the domain wall number \cite{Sikivie:1982qv, Sikivie:2006ni}. We can infer from the equation \eqref{axionmass}, that if $f_a$ is large enough, then the axion can be highly light and stable which, added to the very weak interaction with the rest of matter, makes an excellent DM candidate \cite{Marsh:2015xka, Sikivie:2006ni, chadhaday2021axion}. 

We will focus on the QCD axion models where there are in general three types:
\begin{itemize}
    \item The Peccei-Quinn-Weinberg-Wilczek (PQWW) axion, which introduces one additional complex scalar field only.
    \item The Kim-Shifman-Vainshtein-Zakharov (KSVZ) axion, which introduces heavy quarks as well as the Peccei-Quinn scalar.
    \item The Dine-Fischler-Srednicki-Zhitnitsky (DFSZ) axion, which introduces an additional Higgs field as well as Peccei-Quinn scalar.
\end{itemize}
In these three types, the Lagrangian of each model is taken to be invariant under a global $U(1)$ symmetry, that is spontaneously broken at one scale $f_a$ by the potential, $V(\varphi) = \lambda_{\text{QCD}} \left( \vert \varphi \vert^2 - f_a^2 /2  \right)^2$ where $\varphi$ is the Peccei-Quinn field and takes a vacuum expectation value (\textit{vev}) $\langle \varphi \rangle = f_a / \sqrt{2} $.
In the PQWW model, $f_a \approx 250$ GeV, this scale is accessible to experimental search and given the absence of signals, this axion is excluded by collider experiments.
In KSVZ and DFSZ models the decay constant is a free parameter and can be made large enough such that they are not excluded.  

After the global $U(1)$ symmetry breaking at some scale $f_a$, one angular degree of freedom appears as $\langle \varphi \rangle e^{i\Phi_a / f_a}$. The field $\Phi_a$, is the axion and it is a pseudo Nambu-Goldstone boson of this broken symmetry. 

At the classical level the Lagrangian is invariant under chiral rotation, 
which leads to the shift symmetry of the axion field, $\Phi_a \rightarrow \Phi_a + \text{const}$. But at quantum level non-perturbative physics becomes relevant, e.g. instantons switch on at some particular energy scale $\Lambda_a$ and break the shift symmetry $\Phi_a \rightarrow \Phi_a + \text{const}$, inducing a potential for the axion. However, the potential must respect the residual discrete shift symmetry, $\Phi_a \rightarrow \Phi_a + 2n\pi f_a /N_{\text{DW}}$, for some integer $n$, which remains because the axion is still the angular degree of freedom of a complex field.

The axion potential generated by QCD instantons is,
\begin{equation} \label{pot}
    V_a(\Phi_a) = \Lambda^4_{a} \left[1 - \cos{ \left( \frac{ N_{\text{DW}} \Phi_a }{f_a} \right) } \right] ,
\end{equation}
where $\Lambda_a$ is the non-perturbative physics scale, $N_{\text{DW}}$ is the domain wall number and $f_a$ the PQ symmetry breaking scale. 
If $N_{\text{DW}} > 1$, then there appear domain walls that can quickly dominate the energy density of the early Universe, which is incompatible with standard cosmology and can be avoided if $N_{\text{DW}}$ is taken equal to unity \cite{Sikivie:1982qv, Marsh:2015xka}. 

On the other hand, if we consider only small displacements from the potential minimum $\Phi_a < f_a$, we can expand it as a Taylor series, whose approach to second order is $ V(\Phi_a) \approx \frac{1}{2} \Lambda^4_{a} \Phi_a^2 / f_a^2$. We identified the mass term $\frac{1}{2} m_a^2 \Phi_a^2$, with $m_a^2 = \Lambda_a^4 / f_a^2 $, 

We will adopt as a potential for axion, in subsequent analyses on axions as a dark matter candidate, only the first and second terms of the Taylor series are around the minimum potential, that is, 
\begin{equation}\label{eq:instanton}
    V_a(\Phi_a) = \frac{1}{2} \left( m_a^2 \Phi_a^2 - \frac{1}{12} \frac{m_a^2}{f_a^2}  \Phi_a^4 \right) .
\end{equation}

The axion mass is protected from quantum corrections, since these all break the underlying shift symmetry and must come suppressed by powers of $f_a$. For the same reason, self-interactions and interactions with SM fields are also suppressed by powers of $f_a$. Regarding the self-interactions, we can easily obtain an expression for them by means of an expansion of the cosine potential to higher orders. This renders an axion model with a light (less than meV), weakly interacting, long-lived particle. These properties are protected by a underlying symmetry, so the axion provides a natural candidate to DM model \cite{chadhaday2021axion}.

Some values for the decay constant could be lie around the fundamental scales of particle physics such as Grand Unified Theory (GUT) scale $f_a \sim 10^{16}$\,GeV. Given the lack of knowledge at high-energies\footnote{By high-energies we mean any symmetry breaking scale  $\gtrsim 1$ TeV.}  
structure of particle physics and the difficulties in obtaining well-defined measurements of the initial conditions in inflationary cosmology, there are no strong reasons to prefer any particular value for $f_a$. But usually $f_a \lesssim M_{pl} \sim 10^{19}$ GeV, since it is not obvious how to make a model of such an axion without a full understanding of quantum gravity \cite{Marsh:2015xka, Graham:2013gfa, chadhaday2021axion}. 

A cosmological populations of axions can be produced by various mechanisms, but the main ones are the decay of parent particle, the decay product of topological defect, the thermal population from the radiation bath and the vacuum realignment \cite{Marsh:2015xka}.

In the case of decay of parent particle, a massive particle with $m_X$, is coupled to axion and decays. In all cases $m_X > m_a$ and their decay produces a population of relativistic axions. If the decay occurs after the axions are decoupled from the SM, then they remain relativistic throughout the history of the Universe and become dark radiation \cite{chadhaday2021axion}.

In the case of decay product of topological defect, two scenarios need to be considered: whether the Peccei-Quinn phase transition occurs during or after inflation.

The breaking of global symmetries leads to the formation of topological defects. A broken $U(1)$ creates axion strings and if $N_{DW} > 1$, domain walls appear too \cite{Sikivie:1982qv}. 
If PQ symmetry is broken during inflation, then topological defects and their decay products are diluted by the expansion of the Universe and can be ignored.

In the second stage, after inflation, the PQ symmetry is broken when the radiation temperature drops below $f_a$. The breaking of the global symmetry gives rise to topological defects and the string decay produced axions. The axion field begins oscillating when $m_a \sim H$, these axions are dominated by the low-frequency modes, making them non-relativistic and contributing as CDM to the cosmic energy budget \cite{Marsh:2015xka}. 

If axions are in thermal contact with the standard model radiation, these are created and annihilated during interactions among particles in the primordial soup. The axions established in this way are called thermal axions. 
Initially, axions are in equilibrium with the thermal bath of particles, but later they become decouple at temperature $T_D$.
Thermal axions are relativistic if $T_D > m_a$. Once decoupled the axion population is merely diluted and redshifted by the expansion of the Universe \cite{Sikivie:2006ni}, then the axions become non-relativistic when its temperature is less than $m_a$.
For $f_a > 10^9$ GeV, the thermal axion lifetime exceeds by many orders of magnitude the age of the Universe \cite{Sikivie:2006ni}, but it behaves cosmologically in a manner similar to massive neutrinos, and contributes as hot DM \cite{Marsh:2015xka} suppressing cosmological structure formation. 

In the case of misalignment production, we need to consider the equation of motion for the axion after non-perturbative effects, $\Ddot{\Phi}_a + 3H(t) \dot{\Phi}_a + m^2_a \Phi_a =0 $. It is the equation of a simple harmonic oscillator with $3H(t)$ being time dependent friction. $H(t)$ is  the Hubble parameter.
When $H > m_a$, the axion field is overdamped and it is frozen by Hubble friction, this means that the expansion of the Universe slows the axion field down ($ \dot{\Phi}_a = 0$) and we get a coherent state of axions at rest \cite{Marsh:2015xka}.

The misalignment production of axions is non-thermal and through this mechanism, even very light WIMPs can be Cold Dark Matter \cite{Sikivie:2006ni}.

Axion and ALPs could be located through axion-photon conversion in external electric ($\vec{E}$) or magnetic ($\vec{B}$) fields \cite{Sikivie:1983ip}, described by the Lagrangian
\begin{equation} \label{003}
\mathcal{L}_{A \gamma \gamma} = g_{A \gamma \gamma} \Phi_a \vec{E} \cdot \vec{B},
\end{equation}
where $g_{A \gamma \gamma} $ is the diphoton coupling constant.  
Pseudoscalar-ALPs and scalar-ALPs could be created when a beam of linearly polarized photons propagates in a transverse magnetic field $\vec{B}$. If an optical barrier is placed downstream to the beam, all unconverted photons will be absorbed while ALPs would traverse the optical barrier. By applying a second magnetic field in the regeneration domain beyond the wall, the inverse process can convert the ALPs back into photons, which can be subsequently detected \cite{OSQAR:2015qdv}.
This type of arrangement is called Light-Shining-through-Walls (LSW) experiment and the best current limit has been achieved by the OSQAR (Optical Search for QED Vacuum Birefringence, axions and Photon Regeneration) experiment, with the exclusion limits $\vert g_{A \gamma \gamma} \vert <3.5 \times 10^{-8} \,\textrm{GeV}^{-1}$ at $95\%$ confidence limits, obtained in vacuum for $m_a \lesssim 0.3$ meV \cite{OSQAR:2015qdv}. Other exclusion limits for pseudoscalar and scalar axion-like-particles can be found in \cite{Robilliard:2007bq, Fouche:2008jk, Cameron:1993mr, GammeVT-969:2007pci, OSQAR:2007oyv, Afanasev:2008jt, PVLAS:2007wzd, Ehret:2010mh, ParticleDataGroup:2020ssz}.

In addition to the possible connection to DM, two hints from astro-particle physics strengthen the axion-like particles existence: the anomalous excessive cooling of stars and the anomalous transparency of the Universe to very high energy gamma rays.
The cooling excess can be attributed to ALPs, produced in the hot cores that abandoning the star unimpeded, contributing directly to the energy loss \cite{Giannotti:2015kwo, Giannotti:2015dwa, Giannotti:2017hny}.
The anomalous transparency can be explained if a part of the photons are converted into light spin zero bosons in astrophysical magnetic fields. The ALPs can travel through cosmological distances unhindered, due to their weak coupling to normal matter. A part of such light bosons are in turn reconverted into high-energy photons and could be detected \cite{DeAngelis:2007dqd, Horns:2012fx}.

\subsection{Higgs-like model} \label{extrahiggs}

The Lagrangian density of the Standard Model can be explicitly divided into gauge, fermion, Higgs and Yukawa sectors.
The Higgs part is $ \mathcal{L}_{\varphi} = (D^{\mu} \varphi)^{\dagger} D_{\mu} \varphi - V(\varphi)$, where 
$\varphi = \begin{pmatrix}
        \varphi^{+} \\
        \varphi^{0}
      \end{pmatrix} $  
is a Higgs scalar, transforming as a doublet of $SU(2)$, $\varphi^{\dagger}$ is its  adjoint. $ \varphi^{+}$ and $ \varphi^{0}$ are charge and neutral complex fields and $D_{\mu}$ is the gauge covariant derivative. 
$V(\varphi)$ is the Higgs potential, the combination of $SU(2) \times U(1)$ invariance and renormalizability restricts $V$ to the form 
\begin{equation} \label{pothiggs}
    V(\varphi) = \mu^2 \varphi^{\dagger} \varphi + \lambda (\varphi^{\dagger} \varphi)^2 .
\end{equation}
For $\mu^2 <0$ there will be spontaneous symmetry breaking and the nonzero \textit{vev} of neutral component $\varphi^0$ will generate the $W$ and $Z$ masses. The $\lambda$ term describes a quartic self-interaction $ \lambda (\varphi^{\dagger} \varphi)^2$ of the Higgs field. Vacuum stability requires $\lambda > 0$  \cite{Langacker:2010zza}.

A very useful proposal to explain some particle physics open questions, such as the small mass of the neutrinos \cite{King:2013eh}, the fermionic mixing \cite{Branco:2011iw, Ivanov:2017dad} and dark matter \cite{LopezHonorez:2006gr, Cao:2007rm, Majumdar:2006nt} is the extension of the Higgs sector of the SM, which consists of the introduction of new symmetries, plus the addition of scalar singlets and/or doublets in $\mathcal{L}_{\varphi}$. 
After the break of the electroweak symmetry, the extra scalar fields acquire mass and are known as Higgs-like particles. 

Particularly, dark matter can be explained with inert Higgs scalars, i.e. that do not acquire a \textit{vev}, are stable and cannot decay to SM particles.  The stability is  usually achieved by introducing an extra $Z_2$ discrete symmetry \cite{Espinoza:2018itz}. 

We will adopt for our study a model with an inert Higgs doublet (an equivalent analysis can be done considering a singlet complex scalar field), besides the usual SM one, 

\begin{equation}
   \Phi_h = 
\begin{pmatrix}
    \Phi^{+} \\
    \Phi^{0}    
\end{pmatrix} = \frac{1}{\sqrt{2}}
\begin{pmatrix}
   \Phi_1 + i \Phi_2 \\
   \Phi_3 + i \Phi_4 
\end{pmatrix}  ,
\end{equation}
where $\Phi^{+}$ and $\Phi^{0}$ are the charged and neutral complex components of the field $\Phi_h $, respectively, which can also be expressed in terms of their real parts, $\Phi_i$, $i=1,2,3,4$; whose potential is of the form,
\begin{equation} \label{higgspotential}
    V_h(\Phi_h) = m_h^2 (\Phi^{\dagger}_h \Phi_h) + \frac{\lambda_h}{2} (\Phi^{\dagger}_h \Phi_h)^2,
\end{equation}
where we will choose $m_h^2 > 0$  \cite{LopezHonorez:2006gr,Ivanov:2017dad} 
and none of the components acquire a vacuum expectation value. 
We are assuming that the coupling between the inert doublet and the SM Higgs is very small.
The DM candidate must come from the neutral complex component, $\Phi^0$. It is known from DM experimental searches that this type of matter must be electromagnetically neutral, since the mediator of electromagnetic interaction is the photon and the dark matter is considered to be transparent to light.

In general, the mass of the Higgs-like DM candidates depend largely on the model, for example if they have couplings to the SM fields. But the mass constraints in the experimental search for extra Higgs fields usually lies around the order of GeV. The most recent mass limits for a variety of models with extra neutral Higgs bosons can be found in
\cite{ParticleDataGroup:2020ssz, GAMBIT:2018eea, LHCDarkMatterWorkingGroup:2018ufk}. 

The DM candidate, among the massive states coming from the doublet $\Phi_h $, will be the neutral lightest and stable particle (whose decay is protected by some symmetry).

The consistency of a Higgs-like dark matter model can be checked with the dark matter relic abundance \cite{LHCDarkMatterWorkingGroup:2018ufk}.
According to the WIMP paradigm, the dark matter candidate has weak interactions with the SM particles and was in thermal equilibrium in the early stages of the history of the Universe.
Subsequently, the interaction rate of the DM fell below the Hubble expansion rate causing the freeze-out of the DM \cite{Arcadi:2018pfo}. 

To avoid any confusion, we want to make it clear that in the following, when referring to Higgs particles, we refer to Higgs-like particles ($\Phi_h$), they are different from the SM Higgs doublet ($\varphi$).

Axions, axion-like and Higgs-like particles are excitations of quantum fields, however, the interest in this work is to analyze the behavior of these particles on a cosmological scale, where the DM candidates are treated in a classical way. 
Thus,  we need to make a transition from quantum  to classical theory. This transition can be studied within the framework of an effective action. This topic is described in the next sub-section.

\subsection{Transition from quantum field theory to classical theory.}
\label{sec:transition}
 
Consider a QFT with a Lagrangian density
${\cal L} = {\cal L}_0 + {\cal L}_{\textrm{int}}$. 
For the purpose of this section it is sufficient to consider the example
of one scalar field $\phi(x)$. In the context of a microscopic theory,
it is very important to determine the scattering matrix or $S$-matrix,
since its knowledge allows one to compute observable quantities  like annihilation/scattering
amplitudes for particles including e.g. DM candidates, that can be compared
to observations of indirect/direct DM detection experiments.

We can compute  the $S$-matrix  in terms of the $n$-point Green's functions 
of the theory: 

\begin{equation}
G(x_1,x_2, \ldots x_n) = \bra{\Omega} T\left( \phi(x_1) \cdots \phi(x_n) \right) \ket{\Omega},
\end{equation}
with $\ket{\Omega}$ the vacuum of the interacting theory and $T$ denotes the time-ordering operator.
In the path integral formalism these functions  are encoded in the
generating functional $Z[J]$ through the expression:

\begin{equation}
\begin{split}
&G(x_1,x_2, \ldots x_n) = \\ & (-i)^n \frac{\delta}{\delta J(x_1)} \frac{\delta}{\delta J(x_2)}
\ldots \frac{\delta}{\delta J(x_n)} \,\,  Z[J] \Big|_{J=0}~, 
\end{split}
\end{equation}
where $J(x)$ is an external source and the derivation is functional. The path integral representation of $Z[J]$
is given by

\begin{equation}
Z[J] = N^{-1} \times \int D\phi \exp{i \int d^4x ({\cal L}(\phi(x)) -  \phi(x) J(x))},
\end{equation}
with $N=\int D\phi \exp{i I[\phi]}$ and $I[\phi] = \int d^4x {\cal L}(\phi(x))$.
In cases where there are no 
interactions, the path integral can be evaluated
in closed form taking the free generating functional as:

\begin{equation}
Z_0[J] = \exp{-\frac{i}{2} \int J(x)\Delta_{\textrm{F}}(x-y)J(y)d^4xd^4y},
\end{equation}
where $\Delta_{\textrm{F}}(x-y)$ is the free Feynman propagator.

If interactions are present, then no closed form of the generating functional is known.
However, in this case $Z[J]$ satisfies the differential 
Schwinger-Dyson equation:

\begin{equation}
-i (\Box + m^2) \frac{\delta Z[J]}{\delta J(x)} - {\cal L}^\prime_{\textrm{int}} \left( -i\frac{\delta}{\delta J(x)} \right)
Z[J] = J(x) Z[J],
\end{equation}
where $m$ is the scalar field mass.  The term ${\cal L}^\prime$  denotes the differentiation of ${\cal L}_{\textrm{int}}$
with respect to $\phi$ and evaluated on $\phi \rightarrow -i\frac{\delta}{\delta J(x)}$; the functional differentiation with respect to $J(x)$  acts on $Z[J]$.
The solution to the above equation (up to a normalization factor) can be expressed formally in terms
of the free generating functional as:

\begin{equation}
Z[J] = \exp{\left[ i \int d^4x {\cal L}_{\textrm{int}} \left( -i\frac{\delta}{\delta J(x)} \right) \right]} Z_0[J].
\end{equation}
The exponential in this equation is expressed as a power series in the coupling
constant. This procedure is equivalent to the 
Feynman diagram perturbation theory, thus $Z[J]$ generates all diagrams including disconnected ones. 

There is a generating functional $W[J]$, which generates only connected Feynman diagrams 
(or connected Green's functions).
The connected generating functional $W[J]$ relation to $Z$ is through
exponentiation (switching to normal units to make explicit Planck's constant): 

\begin{equation}
Z[J] = e^{\frac{i}{\hbar} W[J]}.
\end{equation}
 The effective action $\Gamma[{\bar \phi}]$ from $W$ using the Legendre
transformation is 
\begin{equation}
\Gamma[{\bar \phi}] = W[J] - \int d^4x \frac{\delta W[J]}{\delta J(x)} J(x),
\end{equation}
here the following notation has been introduced:
\begin{equation}
\frac{\delta W[J]}{\delta J(x)} \equiv {\bar \phi},
\end{equation}
where ${\bar \phi}$ is called the average (or classical) field.
$\Gamma[{\bar \phi}]$ generates single particle
irreducible connected diagrams. 
This is a simple example of a working analogy where the effective
action is the analog of the Gibbs potential in equilibrium statistical
mechanics in the presence
of coupling to an external source or $J$ reservoir.

The usefulness of the effective action has been shown extensively
in the literature~\cite{Iliopoulos:1974ur}, and we will concentrate on its loop expansion for this work. 
As shown in Ref.~\cite{Jackiw:1974cv}, the effective action 
can be expressed as a series expansion in loops where the $n$-loop term
is proportional to $\hbar^n$:

\begin{equation}\label{effAcc}
\Gamma[{\bar \phi}] = I[{\bar \phi}] + \frac{1}{2} i \hbar \ln \mathrm{det}(i {\cal D}^{-1})
+ \mathcal{O}(\hbar^2),
\end{equation}
where ${\cal D}$ is the propagator for a ``modified'' action, i.e. the action for
the original theory expanded around the average field but keeping only
terms of second and higher order. 
For our present purposes, it suffices to notice that in the limit
$\hbar \rightarrow 0$ the effective action reduces to the tree level
action $I[{\bar \phi}]$, as  expected. Thus, the classical limit of a given theory corresponds to the 0-loop term
in the quantum effective expansion.
It is thus natural to take, for example, the expression 
for the tree level potential of a given particle physics quantum model 
and couple the corresponding classical fields to gravity
as a starting point for an analysis in the context of a cosmological model.

Regarding the quartic parameters, the ones in  the classical action 
will match the zeroth order quantum parameters in the effective expansion. Furthermore, if we assume that quantum corrections are small, the quartic couplings have to lie within the interval $-4\pi < \lambda < 4\pi$, to ensure perturbative unitarity at the quantum level.
Then, the following question arises: Can the physical values of the quartic couplings in the scalar potential, constrained from particle physics, have consequences on the cosmological parameters? 
We will show that in the Higgs-like case, the interval for $\lambda$ implies that this field belongs to the cosmological non self-interacting regime.

The dark matter candidates we have reviewed here, axion, axion-like and Higgs-like, 
are considered as real scalar fields in the classical limit.
However, a more general and appropriate approach in the cosmological framework is to take them as complex scalar fields. 
Since the halos formed by complex scalar field are stationary gravitational solitons known as boson stars, which are stable \cite{Liebling:2012fv} compact objets. On the other hand, the halos in the case of a real scalar field, known as oscillatons \cite{Seidel:1991zh}, are metastable oscillating solutions.

In the case of a Higgs-like particle, it is not entirely correct to say that
the classical limit is a real scalar field; as it happens, in this particular case
this limit is a complex scalar field. 
The transition from a real quantum field to a complex classical field can be understood as follows.

Consider a generic doublet of $SU(2)$, denoted by
\begin{equation}
    H = \begin{pmatrix}
             H^+  \\
             H^0
         \end{pmatrix} = \frac{1}{\sqrt{2}}
         \begin{pmatrix}
            H_1 + i H_2 \\
            H_3 + i H_4
         \end{pmatrix},
\end{equation}
where $H^+$ and $H^0$ are charged and neutral complex components of the Higgs field, respectively. The DM candidate must come from $H^0$, as described in the subsection \ref{extrahiggs}.  
In the quantum scenario, in a first stage, some mechanism at a high energy scale (for instance,  the breaking of a symmetry) will give mass to the components $H_i$, $i=1,2,3,4$ and in principle, the masses of the components of $H^0$ will be equal $m_3 = m_4$, because all these fields form part of the same $SU(2)$ doublet (see Eq. \eqref{higgspotential}).

In a second stage, the electroweak symmetry is spontaneously broken and the SM particles acquire mass. In addition $m_3$ and $m_4$ can acquire radiative corrections, generating an inequality in masses leading to a decay of the heavy particle to the light particle, obtaining only one particle (a real scalar field) as DM candidate.

However, in the classical limit, radiative corrections cannot be detected due to their quantum nature, so the equality $m_3 = m_4$ is preserved, giving us two DM candidates, which can be included as components of a complex scalar field.

\section{A two scalar field model}\label{sec:two_scalar}

We consider two  cosmological scalar fields that contribute to the energy and matter density of the Universe. From this point forward we 
will assume that both are complex and obey the classical field equations, according to the discussion in the previous section. These fields gravitate via minimal coupling given by the action,

\begin{equation}\label{eq:action}
\mathcal{S}=\int d^4 x\sqrt{-g}\left(\frac{c^4}{16\pi G}R+\mathcal{L}_{\Phi_1,\Phi_2}\right),
\end{equation}
where
\begin{equation}\label{eq:Lagrangian}
2\mathcal{L}_{\Phi_1,\Phi_2}=-\nabla^\mu\Phi_1^*\nabla_\mu\Phi_1-\nabla^\mu\Phi_2^*\nabla_\mu\Phi_2-V(\Phi_1,\Phi_2).
\end{equation}

Varying Eq.~(\ref{eq:action}) with respect to the metric $g_{\mu\nu}$ gives 

\begin{equation}\label{eq:einstein}
    R^\mu_\nu-\frac{1}{2} R\delta^\mu_\nu=\frac{8\pi G}{c^4}T^\mu_\nu,
\end{equation}
with

\begin{equation}
\begin{split}
    T^\mu_\nu=&g^{\mu\eta}\partial_{(\eta}\Phi_1^*\partial_{\nu)}\Phi_1+g^{\mu\eta}\partial_{(\eta}\Phi_2^*\partial_{\nu)}\Phi_2\\
    &-\frac{\delta^\mu_\nu}{2}\left[g^{\alpha\beta}\partial_\alpha\Phi^*_1\partial_\beta\Phi_1+g^{\alpha\beta}\partial_\alpha\Phi^*_2\partial_\beta\Phi_2+V(\Phi_1,\Phi_2)\right].
\end{split}
\end{equation}

The variation with respect to the fields $\Phi_1$ and $\Phi_2$ gives the following equations of motion:

\begin{equation}\label{eq:KG1}
    \square \Phi_1-\frac{d V}{d\abs{\Phi_1}^2}\Phi_1=0,
\end{equation}

\begin{equation}\label{eq:KG2}
    \square \Phi_2-\frac{d V}{d\abs{\Phi_2}^2}\Phi_2=0.
\end{equation}

We assume that in addition to the scalar field, we have radiation $r$, baryons $b$ and dark energy $\Lambda$, but these components do not interact with the scalar field. In the homogeneous case, the solution to the Einstein equations (\ref{eq:einstein}), is the Friedman-Lema\^itre metric:
\begin{equation}
    ds^2=-c^2dt^2+a(t)^2(dr^2+r^2d\Omega^2),
\end{equation}
where we have taken the $t=$ constant hyper-surfaces ($k=0$) case, where we consider a flat Universe. The $tt$ component of Eq.~(\ref{eq:einstein}) becomes
\begin{equation}\label{eq:Fridman1}
H^2=\frac{8\pi G}{3c^2}[\rho_r(t)+\rho_b(t)+\rho_\Lambda(t)+\rho_{\Phi_1,\Phi_2}],
\end{equation}
here, $H:=\dot{a}/a$ is the Hubble parameter, $\rho_x$ corresponds to the energy density associated to the energy momentum tensor of $x=r,b,\Lambda$ components and
\begin{equation}
    \rho_{\Phi_1,\Phi_2}=\frac{1}{2c^2}\abs{\partial_t\Phi_1}^2+\frac{1}{2c^2}\abs{\partial_t\Phi_2}^2+\frac{1}{2}V(\Phi_1,\Phi_2).
\end{equation}
In addition to the density of the scalar field, a ``pressure'' term can be defined as
\begin{equation}
    p_{\Phi_1,\Phi_2}=\frac{1}{2c^2}\abs{\partial_t\Phi_1}^2+\frac{1}{2c^2}\abs{\partial_t\Phi_2}^2-\frac{1}{2}V(\Phi_1,\Phi_2),
\end{equation}
and a corresponding equation of state $w$ can be defined as the ratio of density to pressure.

Now, if we concentrate on the case where the potentials are separated for each field $V(\Phi_1,\Phi_2)=V_1(\Phi_1)+V_2(\Phi_2)$, the equations of motion also separate. In this case we have $\rho_{\Phi_1,\Phi_2}=\rho_1+\rho_2$ and $p_{\Phi_1,\Phi_2}=p_1+p_2$, with $\rho_1=\frac{1}{2c^2}\abs{\partial_t\Phi_1}^2+\frac{1}{2c^2}V_1(\Phi_1)$ and similarly for the other density and pressures. In terms of these quantities, the equations of motion for the scalar fields imply the following relations,
\begin{align}
    &\partial_t\rho_1+3H(\rho_1+p_1)=0\label{eq:KG1_p}.\\
    &\partial_t\rho_2+3H(\rho_2+p_2)=0\label{eq:KG2_p}.
\end{align}
Before starting with the technical details on the integration of the coupled complex system of differential equations (\ref{eq:KG1}), (\ref{eq:KG2}) and (\ref{eq:Fridman1}), we need to specify the particular form of the scalar potentials. From now on we will return to natural units.

As discussed in the previous sections, we will consider three scalar potentials, two of them are taken from dark matter scalar fields models hypothetically fundamental and the third, is associated to a scalar field model purely motivated by cosmology.  

As described in the subsection \ref{axion-likesubsec},
QCD non-perturbative effects after the Peccei-Quinn symmetry breaking, at some scale $f_a$, provide a potential for the axion $\Phi_a$. 
A simple choice for this, is the instanton potential in equation \eqref{pot}, which turns out to be a very commonly used potential, if a specific form of self-interaction for the axion is required  \cite{Zhang:2018slz,LinaresCedeno:2020dte,Marsh:2015xka}. 

We will assume that this potential is valid during all the evolution of the Universe. Although axions are described by a real scalar field in the quantum relativistic field theory, at low-energy, axions can be described more simply by a classical non-relativistic effective field theory with a complex scalar field \cite{Zhang:2018slz}. So that we exchange $\Phi_a \to \abs{\Phi_a}$ in \eqref{pot}.

In this analysis we are interested in studying slight deviations from the non-interacting scalar field dark matter model. So, let us consider small displacements of the complex field around the minimum of the potential, $\abs{\Phi_a}\ll f_a$. Then, we can make the expansion of \eqref{pot} as in \eqref{eq:instanton},   $V_a(\Phi_a)=2(m_af_a)^2(\abs{\Phi_a}^2/(2!f_a^2)-\abs{\Phi_a}^4/(4!f_a^4)+\cdots)$. 
Notice that, in this expression we have included an extra 2 factor to be consistent with the Lagrangian of a complex scalar field \eqref{eq:Lagrangian} and for simplicity, we will take just the first and second terms of the expansion, 
\begin{equation}\label{eq:pot_axion}
    V_a(\Phi_a)=m_a^2\abs{\Phi_a}^2-\frac{m_a^2}{12f_a^2}\abs{\Phi_a}^4 .
\end{equation}

We identify the positive self-interaction parameter $\lambda_a / 2 =m_a^2/(12 f_a^2)$, which we will use later.
Equation \eqref{eq:pot_axion}, corresponds to the first potential considered in the subsequent analyses.
Recall that in the low density regime, $\abs{\Phi_a}$ is small compared to $f_a$ and thus, the dynamical behavior is captured by the first terms in the potential \cite{Barranco:2010ib}.

The second scalar potential we will be considering, corresponds to a very massive scalar field that appears in the Higgs-like model. As stated in \ref{sec:transition}, classically the relevant part of the model could manifest itself as a single complex scalar field. 

We assume that this component will be one of the components of dark matter and that it could be modeled in the classical regime by a complex scalar field with the potential\footnote{The Lagrangian of a complex scalar field in QFT usually does not have the overall $2$ factor as in the gravitation references cited here, which coincides with the convention used in this sections (see equation (\ref{eq:Lagrangian})). The $1/2$ term in the $\lambda_h$ term is considered in order to compare directly with the quantum theory, since under the change $\Phi_h\rightarrow\sqrt{2}\Phi_h$ the Lagrangian of the Higgs-like field becomes $\mathcal{L}=-\nabla^\mu\Phi_h^*\nabla_{\mu}\Phi_h-m_h^2\abs{\Phi_h}^2-\lambda_h\abs{\Phi_h}^4$.} \eqref{higgspotential}. 

Typically the mass term $m_h$ is in the GeV region and due to perturbative analysis the self-interaction, lies within $-4\pi<\lambda<4\pi$ as mentioned in section \ref{sec:transition}.

The third single scalar field model introduced to the analysis is one with unrestricted a priori values on mass and self-interaction, in this sense we refer to it as \textit{classical}. We consider this as a (mainly) classical model in which the connection to fundamental physics is somewhat ``free". The parameters of mass and self-interaction are allowed to vary throughout the spectrum of values as long as they are consistent with cosmological and astrophysical observations. We are interested in the (positive) self-interacting case, that has been shown to be necessary according to \cite{Li:2013nal}. The potential for this field is \eqref{eq:potential1}, where $\phi=\Phi_c$, $\mu=m_c$ and $\sigma^2=\frac{1}{2}\lambda_c$.

All three scalar potentials of the single scalar field models
have the same structure, however, we distinguish them by cases given the allowed values of their parameters. The properties of each case are listed in Table \ref{tab:cases}.

\begin{table*}[htbp]
\caption{Single and double scalar field models described in section \ref{sec:two_scalar}. Top: Three single scalar field models with their free parameters and the validity intervals of $m$ and $\lambda$ parameters. The representative cases are the specific values of the parameters explored in this work. Bottom: Three possible double scalar field models with the corresponding combinations at the description. The $\eta$ constraint is referred to the minimum fraction of the energy density of the lightest field at the present ($a=1$) with respect to the total dark matter density. The viability of the models is reported in the last column with two different meanings in the \textit{viability} term for the single models: the (i) column refers to the BBN$+z_\text{eq}$ analysis described on section~\ref{sec:neff}, while the column (ii) denotes the viability from a relic density point of view for the scalar field models reported at the cited works.}
\begin{tabular}{c|c|c|c|c||c|c}  
\toprule
\rule{0pt}{3ex}  \textbf{SINGLE MODEL} & Free & $m$ & $\lambda$&Representative & \multicolumn{2}{c}{\textbf{Viability}} \rule{0pt}{3ex}\\ \cline{6-7} 
\rule{0pt}{3ex} &parameters&&&cases ($m,\, \lambda$)&(i)&(ii) \rule{0pt}{3ex} \\ 
\toprule
\rule{0pt}{3ex}\textbf{Axion} ($\Phi_a$)& $f_a$ & $5.69\left(\frac{10^9\, \mathrm{GeV}}{f_a}\right)\mathrm{meV}$ & $-m_a^2/(6f_a^2)$&($5.7\times 10^{-13}\,\mathrm{eV},-5.4\times 10^{-82}$) &$\times$ &  \checkmark \cite{Marsh:2015xka} \rule{0pt}{3ex}
\\ \midrule
\rule{0pt}{3ex}\textbf{Higgs} ($\Phi_h$)& $m_h,\,\lambda_h$ & $\sim 100\,\mathrm{GeV}$ & $(-4\pi, 4\pi)$& $(100\,\mathrm{GeV},1)$&$\times$ & \checkmark \cite{LHCDarkMatterWorkingGroup:2018ufk} \rule{0pt}{3ex}
\\ \midrule
\rule{0pt}{3ex}\textbf{Classical} ($\Phi_c$)& $m_c,\,\lambda_c$ & $\lesssim1 \,\mathrm{eV}$ &  $>0$& $(3\times10^{-21}\,\text{eV}, 4.2\times10^{-86})$&\checkmark & NA \rule{0pt}{3ex}
\\
\toprule
\toprule 
\rule{0pt}{3ex}\textbf{DOUBLE MODEL} &  \multicolumn{3}{c|}{Description}&$\eta$ constraint & \multicolumn{2}{c}{\textbf{Viability}} \rule{0pt}{3ex}\\ \midrule
\rule{0pt}{3ex}\textbf{I} &  \multicolumn{3}{c|}{Classical + Higgs}& $\gtrsim0.423$ &\multicolumn{2}{c}{\checkmark} \rule{0pt}{3ex}\\ \midrule
\rule{0pt}{3ex}\textbf{II} &  \multicolumn{3}{c|}{Axion + Higgs}& $\times$ &\multicolumn{2}{c}{$\times$} \rule{0pt}{3ex}\\ \midrule
\rule{0pt}{3ex}\textbf{III} &  \multicolumn{3}{c|}{Classical + Axion}& NA\footnote{The analysis of the Model III is an ongoing project and will be reported elsewhere. } & \multicolumn{2}{c}{\checkmark} \rule{0pt}{3ex}\\
\bottomrule
\end{tabular}
\label{tab:cases}
\end{table*}

Starting from these cases, the analysis is carried out on combinations of these. 
However, for simplicity we will explore only two of the three possible combinations, the classical+Higgs and the axion+Higgs. It will be shown that both models have the capability to modify the expansion of the Universe throughout BBN, however it will turn out that the presence of the classical $\lambda>0$ scalar field is required. Therefore the third possible case (classical+axion), together with the first, contain a set of values in their parameters consistent with this analysis, however this model has four free parameters and we will leave the full analysis for a future work.

The first model considers a Higgs-like scalar field in combination with an classical field. The equations of motion for this case have only three free parameters: the mass and self-interaction of the classical and the fraction of it with respect to the Higgs at $a=1$, namely $\eta$ (defined below). The Higgs field is in the weakly self-interacting regime \cite{Suarez:2016eez}, which implies that the field at a homogeneous level behaves similar to the cold dark matter fluid because it always oscillates rapidly. The second model will be the axion+Higgs combination which has one scalar field related free parameter, $f_a$. In the Table \ref{tab:cases} we summarize these models. And as mentioned, the combination axion+classical will not be explored in this work. We will show that the axion field, at most, passes through a matter-like and stiff matter eras and not through the radiation-like era as the $\lambda>0$ case, even so its stiff era may affect expansion sufficiently to influence BBN. 
\\

\section{Cosmological evolution}\label{sec:evol}

Now, with the specific form of the scalar potentials, we are able to continue with the integration of the evolution equations. We take the procedure done by Li et al. in \cite{Li:2013nal} for one complex scalar field. In order to do this, we are going to force both fields to be matter-like at the present time, this means that both of them must be in the fast-oscillation regime, i.e. their complex phase time derivative ($\omega$), must be greater than the Hubble rate: $\omega/H\gg1$. The integration will be made backwards in terms of the variable $a$ (and not $t$), starting at $a=1$, therefore going back in time the fields will come out of the fast oscillating regime but at different times. 

In the standard one-field case, the solution is obtained in two parts given that in the fast oscillation an approximation is needed due to the difficulties of numerical integration. In the two-field case, we must split the domain of $a$ in three, introducing an intermediate section in which one of the fields still oscillates rapidly but the other one is already in transition to the slow oscillation regime. Three sets of differential equations must be taken into account, with adequate initial (or matching) conditions.

Introducing the variables $A_1=\rho_1-p_1$, $A_2=\rho_2-p_2$ and $B_1=m_1^2\partial_t\abs{\Phi_1}^2$, $B_2=m_2^2\partial_t\abs{\Phi_2}^2$ the full system composed of the two (complex) Klein-Gordon equations and the Fridman equation becomesi 

\begin{widetext}
\begin{equation}\label{eq:Fridman2}
    \dot{a}=aH_0\sqrt{\frac{\Omega_r}{a^4}+\frac{\Omega_b}{a^3}+\Omega_\Lambda+\frac{\rho_1}{\rho_{\text{crit}}}+\frac{\rho_2}{\rho_{\text{crit}}}}.
\end{equation}
\begin{align}
&\frac{d\rho_1}{da}=-3\frac{2\rho_1-A_1}{a},\label{eq:rhoa}\\
&\frac{dA_1}{da}=\pm\frac{B_1}{\dot{a}}\sqrt{1+\frac{2\lambda_1}{m_1^4}A_1},\label{eq:Aa} \qquad \text{(If $\lambda_1>0$, take the upper signs. If $\lambda_1<0$ both signs are possible.)}\\
&\frac{dB_1}{da}=-3\frac{B_1}{a} + 2m_1^2\frac{1}{\dot{a}}\left[2(\rho_1-A_1)-\frac{m_1^4}{2\lambda_1}\left(\sqrt{1+\frac{2\lambda_1}{m_1^4}A_1}\mp 1\right)^2\right].\label{eq:Ba}
\end{align}
\end{widetext}

And similarly for the second field. This was showed in the one-field, positive self-interaction case by \cite{Li:2013nal}. The negative self-interaction $\pm$ possibility is explained below. 

Since the equations (\ref{eq:Fridman2}--\ref{eq:Ba}) are solved in terms of $a$, the Friedman equation is purely algebraic. 

Clearly, the initial condition (at $a=1$ since we are integrating backwards) for the density of this components are the values $\Omega_i$ which are well known numbers constrained by observations. In the same way, the density parameter for dark matter, $\Omega_{\text{DM}}$ 
fixes the total energy density $\rho_1+\rho_2$ at $a=1$, therefore we write
\begin{align}\label{eq:k}
    &\rho_1(a=1)=\eta\ \Omega_{\text{DM}}\rho_\text{crit},\\
    &\rho_2(a=1)=(1-\eta)\ \Omega_{\text{DM}}\rho_\text{crit},
\end{align}
for $0\leq \eta\leq1$. This is the only initial condition needed for the fields, since the full system (\ref{eq:rhoa}--\ref{eq:Ba}) reduces to a set of equations in the fast oscillation regime, given below, which as said before, are feasible to solve and ensure that the scalar fields behave like cold dark matter at late times:

\begin{equation}\label{eq:fast}
    p_1=\frac{m_1^4}{9\lambda_1}\left(1\mp\sqrt{1+\frac{3\lambda_1}{m_1^4}\rho_1}\right)^2,
\end{equation}

\begin{equation}\label{eq:p_fast}
    \frac{d\rho_1}{da}=-3\frac{\rho_1+p_1}{a}.
\end{equation}

The same equations (\ref{eq:fast}) and  (\ref{eq:p_fast}) will be applied  to the field $\Phi_2$ exchanging $1\rightarrow2$. The same rule showed near equation (\ref{eq:Aa}) for the $\pm$ signs applies.

This regime was also derived in \cite{Suarez:2016eez} where the solutions where studied in more depth for both the positive and negative self-interaction cases. However, in the case of negative self-interaction, special care must be taken, since there exist two solutions, both with negative pressure. We have chosen the one with increasing pressure, and therefore a candidate for dark matter. This correspond to the so-called normal branch in \cite{Suarez:2016eez}, where in equations (\ref{eq:Aa}), (\ref{eq:Ba}) and (\ref{eq:p_fast}) the upper sign is taken in the $\pm$ expressions.

This division into two branches for the case of negative self-interaction can be extended to the limit of slow oscillations also by means of a simple $\pm$, as indicated in the previous equations. This extension will allow us to obtain the full solution for the negative $\lambda$ case, at least in the normal branch.

Going back in cosmic time, there must be a value for the scale factor at which the fast oscillation regime stops to be valid in one of the fields. At this point, defined by $a_e$, we have to solve the complete system of equations (\ref{eq:rhoa}--\ref{eq:Ba}) given the following matching initial conditions (see \cite{Li:2013nal}): The density and pressure variables are evaluated at $a_e$, thus determining initial values for the density and the variables $A$ in the complete set of equations, while the new variable $B$, must take the value 
\begin{widetext}
\begin{align}
B_1(a_e)=-H(a_e)\frac{\rho_1(a_e)+p_1(a_e)}{\sqrt{1-2\frac{\lambda_1}{m_1^4}(\rho_1(a_e)-p_1(a_e))}}\left(2+\frac{1}{\sqrt{1-3\frac{\lambda_1}{m_1^4}\rho_1(a_e)}}\right),
\end{align}
\end{widetext}
and similarly for $B_2$. 

We solve the equations using a fourth-order Runge-Kutta method. The problem is solved in three parts, as we discussed earlier. We monitor the end of fast oscillation of the fields using the equation for the pulsation in terms of the energy density in the fast oscillation regime \cite{Li:2013nal}, which for the field $\Phi_1$ is,
\begin{equation}\label{eq:omega}
\omega_1:=\frac{d}{dt}\arg(\Phi_1)=m_1\sqrt{\frac{1}{3}-\frac{2}{3}\sqrt{1-\frac{3\lambda_1}{m_1^1}\rho_1}},
\end{equation}

making the switch to the full system when $\omega_1/H\sim10^3$ because according to numerical experimentation, at this point the fast oscillation limit is still valid and at the same time the full system can be solved with computationally reasonable resolution on $a$. A similar value is used in \cite{Li:2013nal} as the threshold.
 
Before proceeding with the cosmological constraints, we show typical solutions for the negative and positive self-interaction single scalar field models in Table \ref{tab:cases}. Then we report representative solutions of the two scalar field models of Table \ref{tab:cases}.

\begin{figure}

{\includegraphics[width = 0.45\textwidth]{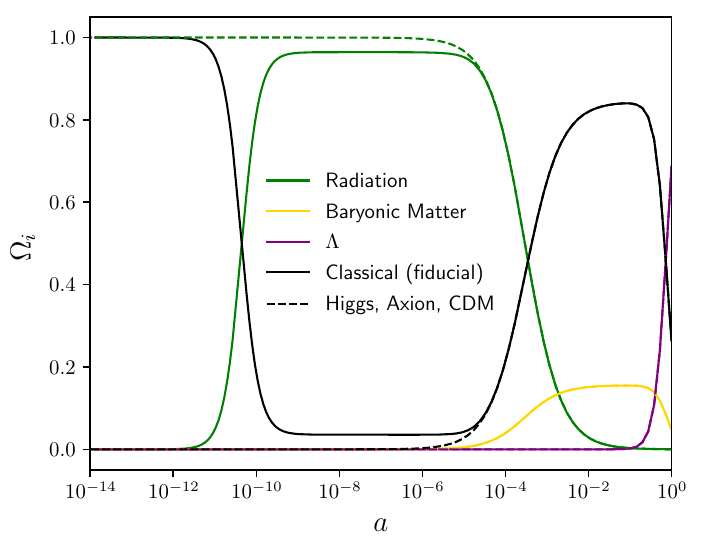}}
\\
{\includegraphics[width = 0.45\textwidth]{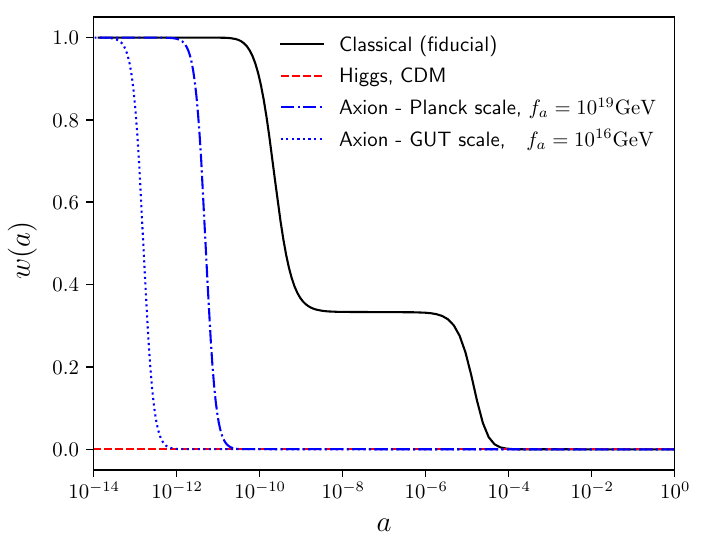}}\\
\caption{Single scalar field representative cases of Table \ref{tab:cases}. Top panel: All solid lines correspond to the classical positive self-interaction fiducial cosmology \cite{Li:2013nal}. The dashed lines are the reference CDM Universe, which happens to coincide in this plot to the Higgs and axion (GUT, Planck) cases. Bottom panel: Equations of state. While the positive self-interaction classical field undergoes three phases, the negative self-interaction case undergoes two and the Higgs field remains indistinguishable from standard cold dark matter.}
\label{fig:planck1}
\end{figure}

In Figure \ref{fig:planck1} we present the plots for the equation of state and the fraction of energy for the following single scalar fields. We have chosen to show some representative cases in each model. For example, for the axion field we take two values of the $f_a$, first the Planck scale $f_a=10^{19}\mathrm{GeV}$ whose mass and self-interaction, according to the formulas in Table \ref{tab:cases}, are $(m_1,\lambda_1)=(5.7\times10^{-13}\mathrm{eV}, -5.40\times10^{-82})$, second the grand unified theory scale (GUT), with $f_a=10^{16}$ corresponding to $(m_1,\lambda_1)=(5.7\times10^{-10}\mathrm{eV}, -5.40\times10^{-70})$.

Then for the Higgs model, we choose a representative case with parameters $(m_1,\lambda_1)=(100\mathrm{GeV},1)$. Actually for the allowed range of $\lambda_h$ the field will always be in the fast oscillation regime. In the extensive study made in \cite{Suarez:2016eez} it was shown that below certain threshold in the parameters, the scalar field will be in the so-called non self-interacting regime. Which can be shown is always the case for the Higgs model, given the big value for the mass and the restriction on $\lambda_h$. In other words, taking different (allowed) values for $\lambda_h$ and even lowering 9 orders (or raising any order) $m_h$ gives indistinguishable solutions between them and also indistinguishable from $\Lambda$CDM. In Figure \ref{fig:planck1} we use the same red line in the top panel, to describe the equation of state of the Higgs as well as the one of the fluid standard CDM. 

We note that for the axion fields, the transition from stiff $w=1$ to matter-like $w=0$, occurs later as the scale $f_a$ increases. However, for the Planck scale axion this is still not enough to noticeably change the density fractions of the components of the Universe in this particular range for $a$ between $10^{-14}$ and $1$, this is the reason that in the plot for $\Omega_x$, the bottom panel in Figure \ref{fig:planck1}, we use the same dashed black line to describe standard CDM as well as the axion and the Higgs models. Important differences in the density parameters (including at BBN), are obtained for $f_a$ above the Planck scale; e.g. for a mass term of the order $m_a\sim10^{-20}\text{eV}$ whose scale of symmetry breaking is $f_a\sim10^{26}$.

Other reference case of interest is the fiducial model in \cite{Li:2013nal}, obtained as a classical positive self-interaction scalar field that satisfy some cosmological constraints involving BBN value of the effective number of neutrinos and the behavior of the field at $z_{\mathrm{eq}}$. This is a model with parameters $(m_1,\lambda_1)=(3\times10^{-21}\text{eV}, 4.2\times10^{-86})$. 

\begin{figure*}
\subfloat[Two scalar field model I. Density fractions for $\eta=0.25$]{\includegraphics[width = 0.45\textwidth]{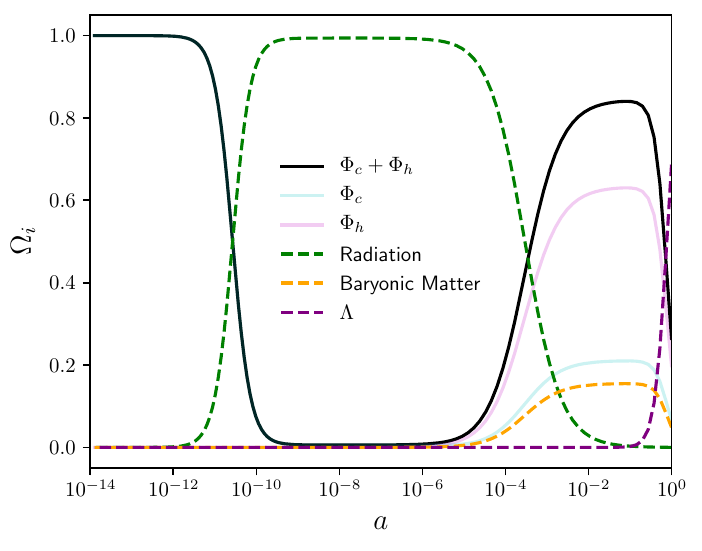}}
\subfloat[Two scalar field model I. Density fractions for $\eta=0.75$]{\includegraphics[width = 0.45\textwidth]{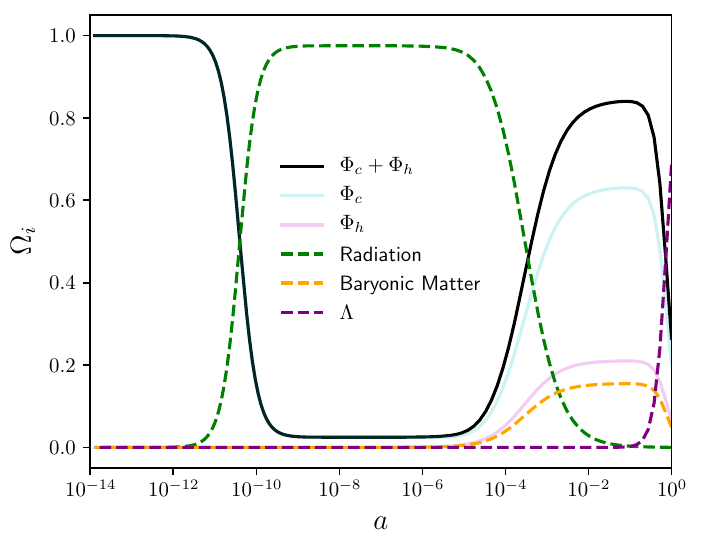}}\\

\subfloat[Two scalar field model II. Density fractions for $\eta=0.25$]{\includegraphics[width = 0.45\textwidth]{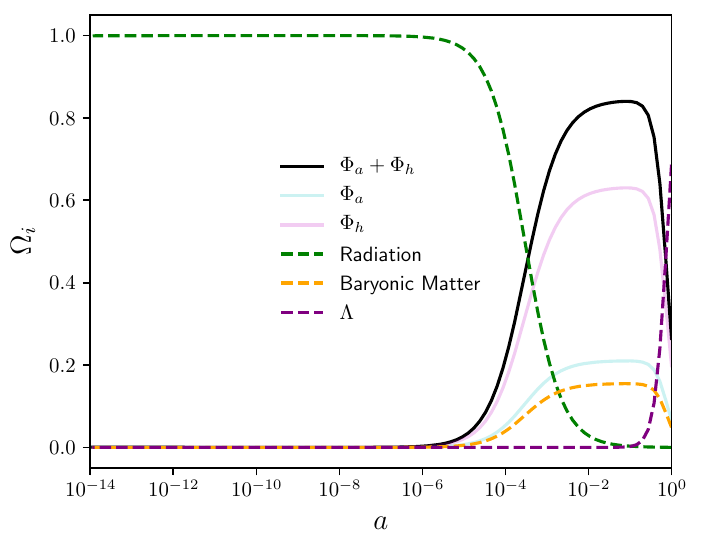}}
\subfloat[Two scalar field model II. Density fractions for $\eta=0.75$]{\includegraphics[width = 0.45\textwidth]{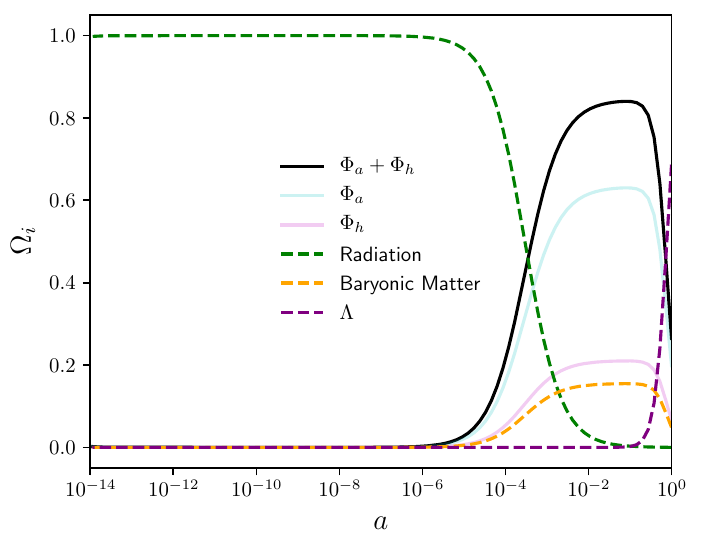}}
\caption{Evolution of the density parameters of the Universe. All solid lines correspond to the scalar field dark matter model with two components and the dashed lines represent the rest of the density contributions. Top panel: Two scalar field model I (Classical + Higgs). Bottom panel: Two scalar field model II (axion+Higgs).}
\label{fig:planck2}
\end{figure*}

In the top panel of Figure~\ref{fig:planck2} we present the fraction of energy for the two scalar field model constituted by the classical and the Higgs fields, that we name  Model I, with two different contributions of each scalar field. The scalar field parameters for the representative case plotted in the top panel of Figure~\ref{fig:planck2} are given in Table~\ref{tab:cases}. We can see the evolution of the dark matter density of the Model case I with a solid line and the contribution of the individual scalar fields with translucent lines. All the other contributions are plotted in dashed lines. Finally, at the bottom panel of Figure~\ref{fig:planck2} is plotted the fraction of energy of the Model II, assembled with an axion field and a Higgs field with the values of $m$ and $\lambda$ provided on Table~\ref{tab:cases} for the representative cases.

\section{Observational Constraints}\label{sec:obsevational}

\subsection{Variability on the cosmic ladder}\label{sec:ladder}


In this section we compute the variation of our model using:
\begin{itemize}
    \item I: Classical+Higgs,
    \item II: Axion+Higgs.
    \end{itemize}
see Table~\ref{tab:cases}. In order to perform such analysis we considered the following: 
\begin{enumerate}
\item An early $H_0$ prior: using the cosmological model $\Lambda$CDM and the Planck Collaboration \cite{Planck:2018vyg} data with the corresponding $H_0=67.04\pm 0.5$ km\,s$^{-1}$ Mpc$^{-1}$. We denote this prior as PL18.
\item A late $H_0$ prior: from the measurement of the Cepheid amplitudes at late times with the corresponding $H_0=74.03\pm 1.42$ km\,s$^{-1}$ Mpc$^{-1}$ \cite{Riess:2019cxk}. We denote this prior as R19.
\end{enumerate}
Following this recipe would give us a percentage rate of the differences between the models described above and the scale factor at which these deviations take place.

\begin{figure*}
\subfloat[Variability of the Model I with $\eta=0.25$]  {\includegraphics[width = 0.45\textwidth]{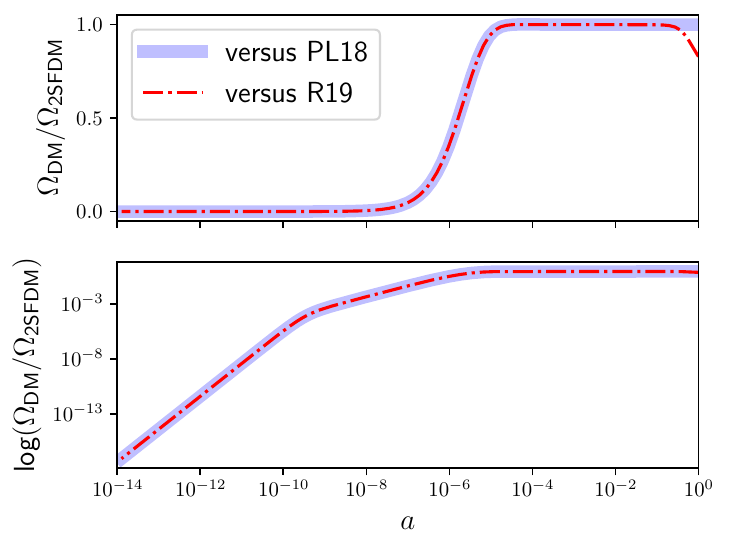}}
\subfloat[Variability of the Model I with $\eta=0.75$]{\includegraphics[width = 0.45\textwidth]{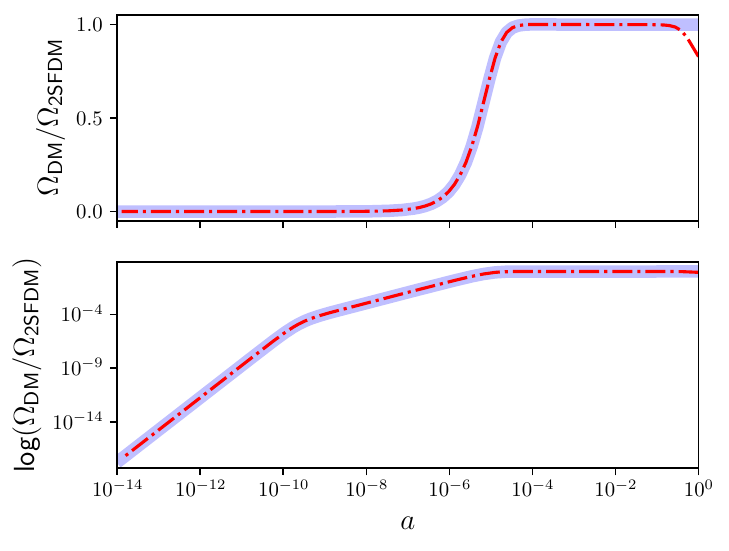}}\\
\subfloat[Variability of the Model II with $\eta=0.25$]{\includegraphics[width = 0.45\textwidth]{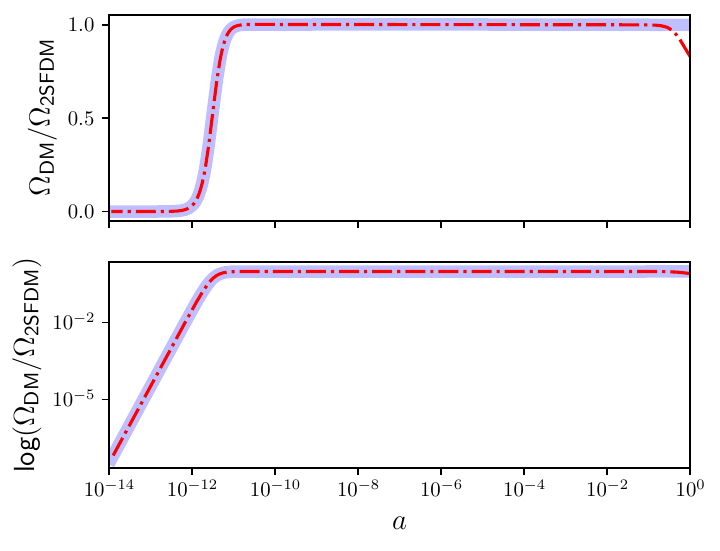}}   
\subfloat[Variability of the Model II with $\eta=0.75$]{\includegraphics[width = 0.45\textwidth]{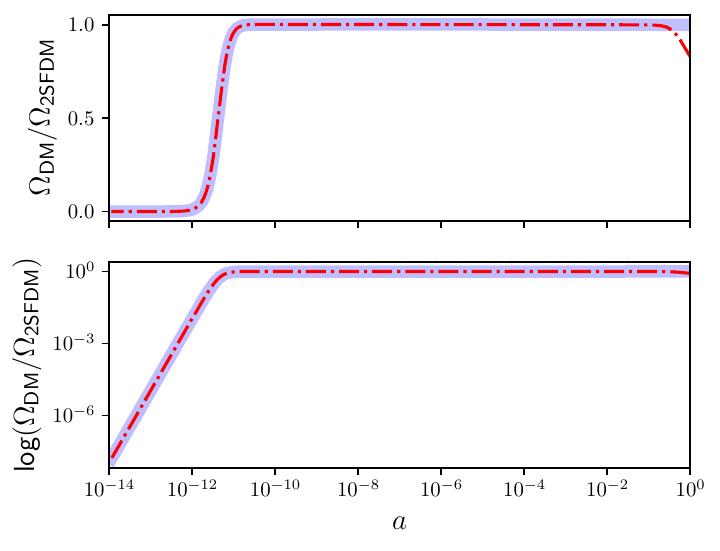}} \\
\caption{Variability of our model, labeled as 2SFDM, with respect to Planck  2018 (PL18) and Riess et al (R19) $H_0$ priors in (a) the case I (Classical+Higgs) for $\eta=0.25$ and (b) for $\eta=0.75$. (c) The variability in the case II (axion+Higgs) for $\eta=0.25$ and (d) for $\eta=0.75$. The values for the free model parameters are in Table~\ref{tab:cases}}
\label{fig:var_contentk}
\end{figure*}

\begin{figure}
{\includegraphics[width = 0.45\textwidth]{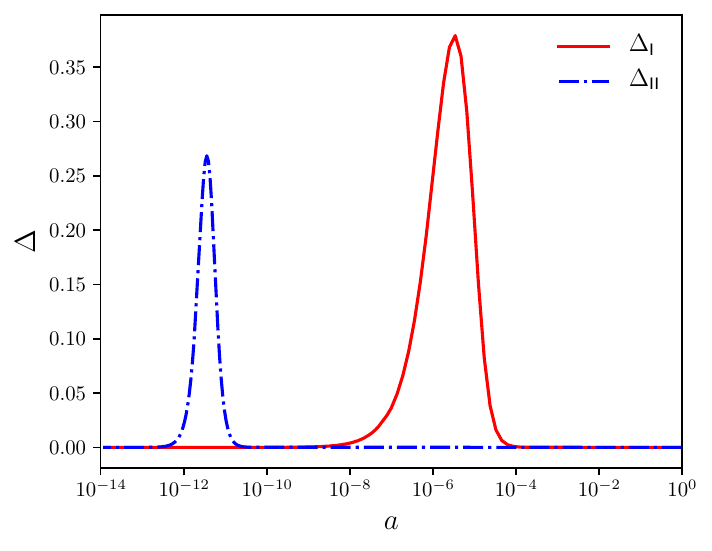}}
\caption{Quantification (Eq.~\ref{eq:delta}) of the shift on the slope of the variability (see Figure \ref{fig:var_contentk}) of the models I and II. The maximum relative difference in the density fraction between $\eta=0.25$ and $\eta=0.75$ for the Classical + Higgs model computed with R19 relative to the density fraction of the two scalar field model, occurs at $a\sim 10^{-6}$ (red solid line); while for the axion + Higgs model this occurs earlier, at $a\sim 10^{-12}$ (blue dashed line). This behavior is the same for the density fraction when PL18 is used and shows that the model I is more sensitive to the current energy density fraction than the model II.}
\label{fig:var_contentk2}
\end{figure}

We solved the system of equations (\ref{eq:Fridman2}-\ref{eq:Ba}) as explained before and compute the corresponding cosmological evolution for the Classical, axion--like and Higgs--like scalar fields at different contributions of each one, characterized by the $\eta$  parameter, for each case analyzed and with the two different $H_0$ priors mentioned above. Then we compare this evolution of the dark matter fractional densities $\Omega_\text{DM}$ with that of our model, noticing that the main difference, characterized by the slope in the upper plots in Figure~\ref{fig:var_contentk}, has a dependence on the $\eta$ parameter. That is, the larger the parameter, the more the slope shifts to the right. This movement can be quantified if we compute 
\begin{equation}
    \Delta = \frac{\Omega_\text{DM} (\eta=0.25)-\Omega_\text{DM} (\eta=0.75)}{\Omega_\text{2SFDM}},
    \label{eq:delta}
\end{equation}
where $\Omega_\text{2SFDM}$ is the density fraction of the two scalar fields and $\Omega_\text{DM}$ is the density fraction computed with PL18 and R19, i.e, $\Delta_\text{I}$ refers to the case I and $\Delta_\text{II}$ corresponds to the case II. As we can see in Figure~\ref{fig:var_contentk2}, the quantification on the shift due to the difference of the $\eta$ parameter is bigger in the model I compared with model II, showing that the model constituted by classical and Higgs fields is more sensitive to the current energy density fraction of a lighter scalar field than the model made of axion and Higgs scalar fields. Furthermore, we can notice that, without taking into account the variation on the $\eta$ parameter, the differences between the values of the dark matter density fraction with PL18 and R19 priors and the density fraction $\Omega_\text{2SFDM}$ of our model occur in an earlier Universe on the model II than in the model I. See, e.g. Figure \ref{fig:var_contentk} (a) and (c).

\subsection{Constraints from \texorpdfstring{$N_{\mathrm{eff}}$ and $z_\mathrm{eq}$}{N and z}}\label{sec:neff}

Among the parameters that determine the production of light elements at BBN we have the expansion rate $H$. This is a period where every component other than radiation is subdominant, therefore the presence of extra relativistic degrees of freedom, beyond the Standard Model implies a modification to $H$ with respect to its $\Lambda$CDM profile. This can be quantified inside the effective number of neutrino species $N_{\mathrm{eff}}$ as a contribution to the $\Lambda$CDM value $N_{\mathrm{eff}}^0$ through a parameter known as number of equivalent neutrinos $\Delta N_\nu$, although its source does not necessarily come from a neutrino. It is defined by
\begin{equation}\label{eq:deltaN_def}
    \Delta N_\nu=\frac{\rho_\xi}{\rho_\nu},
\end{equation}
where $\rho_\nu$ is the energy density of the standard model neutrino (per neutrino specie) and $\rho_\xi$ is the energy density of the additional relativistic fields in consideration, this contribution could correspond to the positive self-interaction (classical) scalar field or to the negative self-interaction axion field for those cases when the energy contribution is important in order to modify $H$, that is, when they behave as radiation and/or stiff matter during BBN. With the previous definition, the total radiation energy density divided by the photon energy density, $\rho_\gamma$, is
\begin{eqnarray}
    \frac{\rho_r}{\rho_\gamma}=1+\frac{\rho_\nu}{\rho_\gamma}\left(3+\Delta N_\nu\right).
\end{eqnarray}
If it is assumed that neutrinos are completely decoupled from the electromagnetic plasma at the electron-positron annihilation, then the temperature of the photons increases with respect to that of the neutrinos by $(T_\nu/T_\gamma)^3=4/11$. Now, the density ratio $\rho_\nu/\rho_\gamma=7/8(T_\nu/T_\gamma)^{4}$, implies that
\begin{equation}\label{eq:rhogamma}
        \frac{\rho_r}{\rho_\gamma}=1+\frac{7}{8}\left(\frac{4}{11}\right)^{4/3}N_{\mathrm{eff}},
\end{equation}
with 
\begin{equation}\label{eq:Neff_def}
N_\mathrm{eff}=N^0_\mathrm{eff}\left(1+\frac{\Delta N_\nu}{3}\right); \ \ N_\mathrm{eff}^0=3\left[\frac{11}{4}\left(\frac{T_\nu}{T_\gamma}\right)^3\right]^{4/3}.
\end{equation}

Where in this case $N^0_\mathrm{eff}=3$. However if it is not assumed that the neutrinos are completely decoupled when the electron-positron pairs annihilate, then $N^0_\mathrm{eff}=3.046$ \cite{Mangano:2005cc}. 

The total $N_\mathrm{eff}$ enters through $H$ to the equations that determine the primordial light element abundances (solved by BBN codes) and if for example the lepton asymmetry is neglected, then the BBN primordial abundances  can be confronted with astronomical observations of the abundances of (mainly) deuterium D \cite{Pettini:2012ph} and the isotope $^4\mathrm{He}$ \cite{Izotov:2013waa}. These constraints on the observed elements can be traduced in constraints over $N_\mathrm{eff}$ as well as $\Omega_b$ \cite{Steigman:2012ve,Nollett:2014lwa}.

In 2015 reference \cite{Nollett:2014lwa}, obtained $N_\mathrm{eff}=3.56\pm0.23$ or

\begin{equation}\label{eq:deltaN}
    \Delta N_\nu=0.5\pm0.23.
\end{equation}

This value certainly excludes the possibility of a new neutrino as well as the standard $N^0_\mathrm{eff}$ case. Nevertheless the parameters of a complex scalar field with positive $\lambda$ can be constrained to be consistent with this measurement as showed by Li et al. \cite{Li:2013nal, Li:2016mmc} if a time dependent $\Delta N_\nu(a)$ is assumed rather than a relatively late time fixed value. The constraint (\ref{eq:deltaN}) is applied through BBN, between the neutron to proton freeze-out and the first nuclei production, at $a_\mathrm{n/p}$ and $a_\mathrm{nuc}$ respectively. 

In our numerical analysis, if $\Phi_1$ is not subdominant at BBN, then the constraint (\ref{eq:deltaN}) is implemented with the formula,

\begin{equation}
\label{eq:Neffcode}
\begin{split}
    &N_{\mathrm{eff}}=\frac{N_{\mathrm{eff}}^0}{2}
    \left(1+\frac{\Omega_1}{\Omega_r}\right.\\
    &\left.+\sqrt{\left(1+\frac{\Omega_1}{\Omega_r}\right)^2+\frac{\Omega_1}{\Omega_r}\frac{32}{7}\left(\frac{11}{4}\right)^{4/3}\frac{1}{N_{\mathrm{eff}}^0}}\right),
    \end{split}
\end{equation}

which is a result of inserting $\rho_\gamma$ from (\ref{eq:rhogamma}) into (\ref{eq:Neff_def}) along with the definition (\ref{eq:deltaN_def}) and $\rho_\nu=\Omega_r-\Omega_\gamma$, notice that in this expression, $\Omega_r$ contains the $\gamma$ and $\nu$ contributions only, which are evolved separately from $\Phi$ in the code.

Additional to the BBN constraints discussed so far, there is a need to make the relativistic and stiff matter scalar field solutions reach a matterlike behavior in $w$ at the latest in the matter-radiation equality $z_\mathrm{eq}\approx3365$. This condition is imposed in the code by setting $w(z_\mathrm{eq})<0.001$.

The results of this BBN+$z_\mathrm{eq}$ analysis for the single $\lambda_1>0$ scalar field, was reported first by Li et al. in \cite{Li:2013nal} and later an update was made within their work \cite{Li:2016mmc}. We recover their result: 

\begin{align}\label{eq:constr_lambdap}
    &m\gtrsim5\times10^{-21}\ \mathrm{eV},\quad \text{(single $\lambda>0$)} \\
    8\times10^{-4}&\ \mathrm{eV}^{-4}\lesssim\frac{\lambda_1}{m_1^4}\lesssim10^{-2}\ \mathrm{eV}^{-4}.
    \label{eq:constr_lambdap2}
\end{align}

If we repeat this analysis now including the single scalar axion case, we should be able to obtain a constraint on the single parameter $f_a$ particularly for the cases with big values of this parameter, which as shown in the previous section, are the models that affect expansion the most. It should be mentioned that the general $\lambda<0$ case cannot be solved in all the cases, particularly in those where the slow oscillation regime appears closer to $a=1$ and the square root arguments in (\ref{eq:Aa}) and (\ref{eq:Ba}) become negative at certain point $a_i$ which corresponds to a place where the scalar field ``turns on'' \cite{Suarez:2016eez}. Luckily, numerical experimentation on solutions for the axion field (where the mass and self-interaction have a specific dependence on $f_a$) shows that this is never the case and no discontinuities in the Einstein equations appear. 

However, the situation occurs when the stiff matter stage of the axion affects $N_\mathrm{eff}(a)$ very drastically, and not in the ``stepped" way in which it happens for the $\lambda>0$ case. If the limits are kept to $1\sigma$ in equation (\ref{eq:deltaN}) then there is no value of $f_a$ for which $N_\mathrm{eff}(a)$ is kept inside these limits, not even at $2\sigma$. It happens that if the $N_\mathrm{eff}(a)$ enters into the limits (\ref{eq:deltaN}) in $a_\mathrm{n/p}$ at the beginning of nucleosynthesis, then it no longer enters at the end of it, at $ a_\mathrm{nuc}$, and vice versa.

Therefore, the single axion model is discarded in relation to this cosmological constraint.

It is possible to repeat this analysis for the two scalar field cases. We are interested in exploring Model I and Model II (Table \ref{tab:cases}). Both of them include the Higgs-like field, which as has been said is similar to CDM fluid regardless of the specific values that $m_h$ and $\lambda_h$ assume. Therefore, in Model I we have a three parameter model and in Model II we have just two parameters.

\begin{itemize}
\item Model I.
We fix the value of $\eta$, (i.e. the fraction of the energy density of $\Phi_c$ at $a=1$ with respect to total dark matter density, (\ref{eq:k})), and explore the existence of possible values of $m_1$ and $\lambda_1$ consistent with the $1\sigma$ BBN+$z_{\mathrm{eq}}$ analysis. The case $\eta=1$ coincides with the single case constraints in (\ref{eq:constr_lambdap},\ref{eq:constr_lambdap2}). If we begin to decrease the value of $\eta$, the range of the parameters consistent with the constraint also decrease in size, as shown in Figure \ref{fig:Neff}, until a critical value is reached, after which no value is allowed. This constraint on $\eta$, gives
\begin{equation}
    \eta\gtrsim0.423\ .
\end{equation}
That is, an upper bound of $\sim 58\%$ for the Higgs (or $w=0$ fluid) component can be considered in order to be consistent with these constraints. In the critical case, where $\eta$ takes values near 0.423, we have that the $(m,\lambda/m^4)$ parameter space narrows to the values $m\gtrsim 2\times10^{-21}\mathrm{eV}$ and $\lambda/m^4\sim3\times10^{-2}\mathrm{eV}^{-1}$.

\begin{figure*}
\subfloat{\includegraphics[angle=0,width=0.5\textwidth]{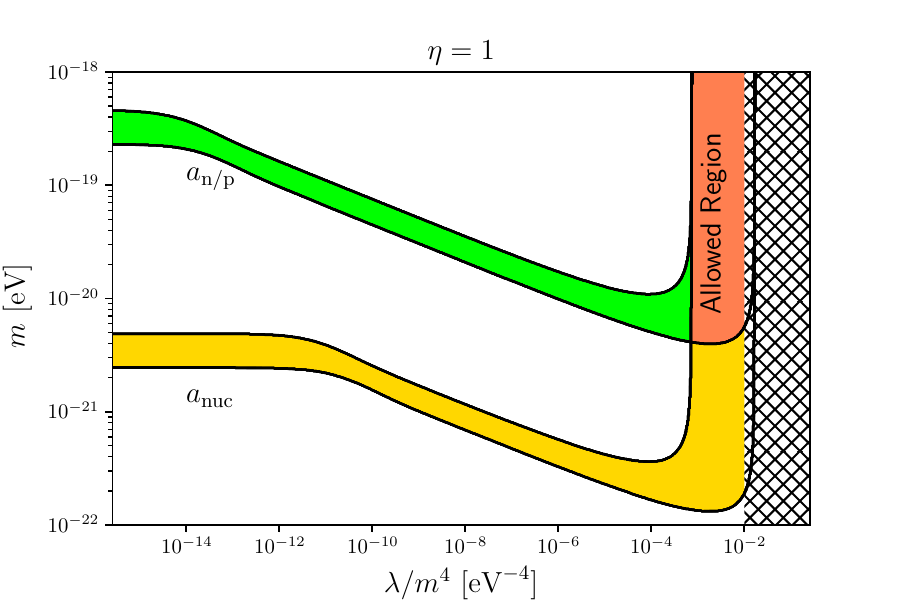}} 
\subfloat{\includegraphics[angle=0,width=0.5\textwidth]{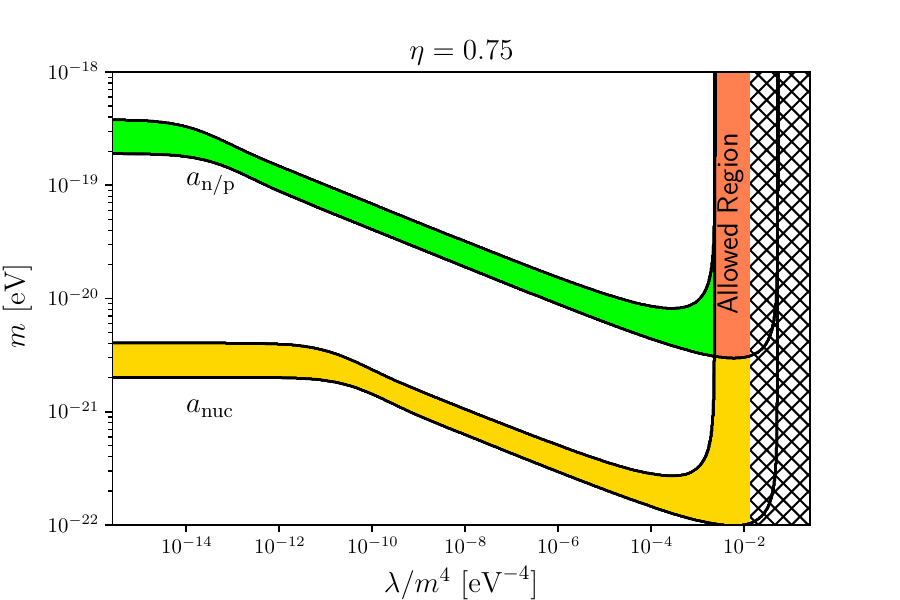}}\\
\subfloat{\includegraphics[angle=0,width=0.5\textwidth]{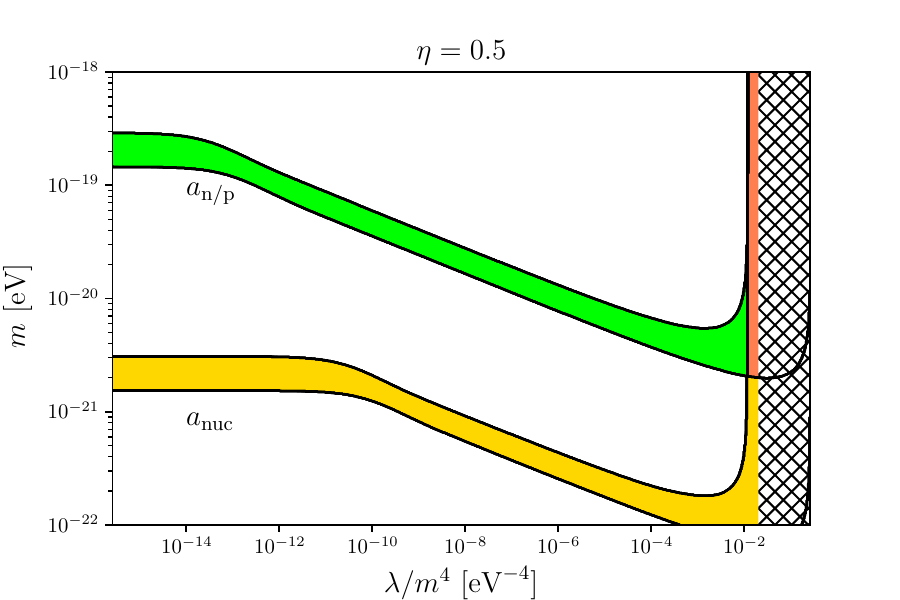}}
\subfloat{\includegraphics[angle=0,width=0.5\textwidth]{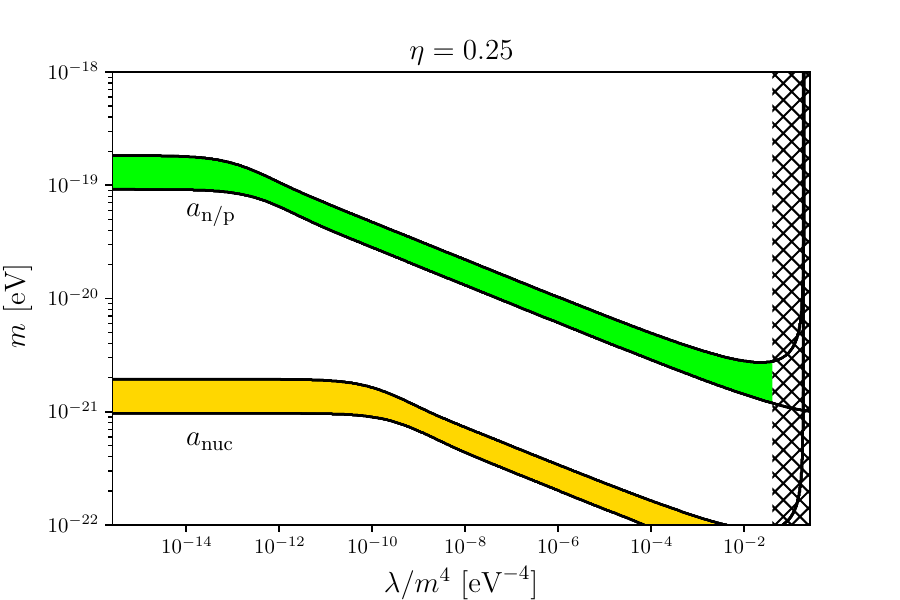}}
\caption{constraints from $z_\mathrm{eq}$ and $N_{\mathrm{eff}}$ within $1\sigma$ for the two scalar field Model I. $\eta$ is the fraction of the classical field with respect to the total dark matter components. The crosshatched region that appears on the right side of all figures, represents the values of the scalar field parameters not allowed by the $z_{\mathrm{eq}}$ constraint. The green and yellow bands are the allowed regions from the $N_\mathrm{eff}$ constraint, (\ref{eq:deltaN}), at $a_\mathrm{n/p}$ and $a_\mathrm{nuc}$ respectively. The red band is the region of the parameter space that is consistent with both the $z_\mathrm{eq}$ and $N_\mathrm{eff}$, throughout BBN, constraints. 
}
\label{fig:Neff}
\end{figure*}

\item Model II.
In this simpler case, a joint analysis over $\eta$ and $f_a$ can be made.
We find that no $1-\eta$ ratio of the Higgs field is capable of smoothing the $N_\mathrm{eff}(a)$ evolution dictated by the axion, during BBN. And since the Higgs field has a contribution to $N_\mathrm{eff}$ of 0 with respect to $N_\mathrm{eff}^0$, this two fields case (like the single axion case), is discarded in the sense that there is no set of parameters such that (\ref{eq:deltaN}) is satisfied. Relaxing the constraint to $2\sigma$ in (\ref{eq:deltaN}) no allowed values are found either.

\item Model III. 
The classical ($\lambda>0$) + axion case has 4 relevant free parameters and a higher complexity. It is found that for all axions with $f_a<10^{19}\mathrm{GeV}$, for which the equation of state is 0 before the start of BBN, as showed in section \ref{sec:evol}, the axion scalar field behaves effectively as a $w=0$ fluid for the purposes of this restriction, and therefore the same restriction as for case I would apply here. A complete analysis of this case will be reported elsewhere.

\end{itemize}

\section{Conclusions}\label{sec:conclusions}
 
 
 In this work we presented a straightforward analysis
 to incorporate scalar fields derived from models coming from physics beyond the Standard Model of particle physics (BSM) in the cosmological evolution. 
 The usual Cosmological models that incorporate scalar fields to describe the dark matter component of the Universe, have been successful in building  a serious alternative to the well known CDM model. These type of  proposals consider a scalar field that does not interact in any way, except via the gravitational interaction, with the rest of the matter in the Universe; we have denoted these fields as {\it classical}.
 
 The combination of some  scalar fields coming from BSM with the classical scalar field proposal, demands a clear description and discussion of the interpretation of the transition from a quantum field theory to a 
 wave function satisfying the Einstein-Klein-Gordon system of equations. 
In order to do this transition we used the effective action perturbative expansion, where we identified  the 0-th order term with the classical field, which we then reparametrized as a complex scalar field. 
 
The BSM fields that we analyzed were the Higgs-like and axion-like fields (clearly the SM Higgs boson itself, being the mass mediator, cannot be used to describe the dark matter), and included them along with the classical one considering that the dark matter is composed of two such fields.  Then, both of these fields would contribute to the dark matter relic density observed today, i.e. $0.26$, and we explore  which proportions of each field today are consistent with BBN at the early Universe. 
 
To start, we considered the case of only one BSM field, taken as a complex classical field as explained in  sub-section \ref{sec:transition}.  We found that in the case of a single Higgs-like DM, it is not possible to increase $N_\mathrm{eff}$ in a significant way during BBN, notice that this is the same case as in $\Lambda$CDM. On the other hand, the axion and axion-like fields increase $N_\mathrm{eff}$ abruptly without the possibility of satisfying the constraint during all the period of BBN.

Neither is it possible to find allowed values consistent with this constraint for the BSM parameters in the case where both fields are combined in any proportion. On the other hand, 
to produce a cosmological model that remains consistent with constraints satisfied by the single classical scalar field, we can consider up to $58\%$ of $\Omega_\mathrm{DM}$ to be a Higgs-like field if the remaining $42\%$ is the classical one. The combination of a classical with an axion-like field turns out to have four free parameters that prevent us from performing a brief survey of the parameters and make a full analysis in the lines of the present work. However, we can say in advance that a combination of a classical field together with an axion or axion-like field, will have a set of parameters for the scalar field where this restriction is satisfied, specifically for $f_a<10^{19}\,\mathrm{GeV}$.
 
We want to stress the results regarding the Higgs-like scalar field. The searches on direct  \cite{2013EPJC...73.2648B, 2015arXiv150400820D, 2016EPJC...76...25A, 2017PhRvL.119r1302C} and indirect \cite{2011ICRC....5..141D, 2015JCAP...09..008F,2020NuPhA.99621712Y} detection of dark matter usually take into account one DM candidate, which comprises 100\% of the relic density.  Our result opens the possibility to 
 take into account in the direct and indirect searches more than one candidate to DM, which contribute to the relic density in different proportions, and thus modify the expected fluxes in the experimental analysis. This fact has to be taken into account in the design of the experiments and in the interpretation of their results. 
 
 In any case, 
 according to BBN and $z_\mathrm{eq}$ analysis, a large part of $\Omega_\mathrm{DM}$ in our two field models, is required to be the classical complex scalar field, which has zero interaction with the rest of the matter, beyond the gravitational one. In order to understand more of its properties, different types of experiments have to be developed. For instance, the distribution of the complex scalar field in the vicinity of a black hole, so called black hole wigs \cite{Barranco:2012qs}, has a very particular density distribution which, in turn, affects the dynamics of light and observable matter the vicinity in a characteristic way. It is important to look, as discussed and done in \cite{Nunez:2015vld} for instance, for possible observable (gravitational) consequences of one type or another of dark matter model, in order to be able to discard or make more robust a given proposition for describing that quarter of the total density of the Universe that we call dark matter.

\begin{acknowledgments}
This work was partially supported  by the CONACyT Network Projects No. 376127 ``Sombras, lentes y ondas gravitatorias generadas por objetos compactos astrof\'isicos", and No. 304001 ``Estudio de campos escalares con aplicaciones en cosmolog\'ia y astrof\'isica".
LEGL, BCM and VJ acknowledge financial support from CONACyT graduate grants program 786444, 660893 and 831066 respectively.
CE-R is supported by DGAPA-PAPIIT UNAM Project TA100122 and acknowledges the Royal Astronomical Society as FRAS 10147. CE, LG and MM acknowledge support by DGAPA-PAPIIT-UNAM project IN109321. CE acknowledges the support of CONACyT C\'atedra 341. The authors are grateful for the computational time granted by ICN-UNAM in the ``Tochtli'' Cluster.
\end{acknowledgments}

\bibliographystyle{apsrev4-2}
\bibliography{paperSFPC}

\end{document}